\documentclass[preprint,12pt,authoryear]{elsarticle}

\usepackage{amssymb}

\usepackage{fancyhdr}
\usepackage{pdflscape}
\usepackage{booktabs}
\usepackage{longtable}
\usepackage{caption}
\usepackage{subcaption}
\usepackage{hyperref}
\hypersetup{
    colorlinks=true, 
    urlcolor=blue,
    }
\usepackage{float}

\journal{Computers in Biology and Medicine}

\begin{document}

\begin{frontmatter}



\title{Automated neuroradiological support systems for multiple cerebrovascular disease markers -  A systematic review and meta-analysis}

\author[inst1,inst2]{Jesse Phitidis\corref{cor1}}
\ead{j.phitidis@ed.ac.uk}
\cortext[cor1]{Corresponding author}
\author[inst2,inst3]{Alison Q O'Neil}
\author[inst1]{William N Whiteley}
\author[inst4,inst5]{Beatrice Alex}
\author[inst1,inst6]{Joanna M. Wardlaw}
\author[inst7]{Miguel O. Bernabeu}
\author[inst1,inst6]{Maria Vald\'{e}s Hern\'{a}ndez}

\affiliation[inst1]{organization={Centre for Clinical Brain Sciences, University of Edinburgh},
            addressline={49 Little France Crescent}, 
            city={Edinburgh},
            postcode={EH164SB}, 
            country={United Kingdom}}

\affiliation[inst2]{organization={Canon Medical Research Europe},
            addressline={Bonnington Bond, 2 Anderson Place}, 
            city={Edinburgh},
            postcode={EH65NP}, 
            country={United Kingdom}}

\affiliation[inst3]{organization={School of Engineering, University of Edinburgh},
            addressline={Sanderson Building}, 
            city={Edinburgh},
            postcode={EH93FB}, 
            country={United Kingdom}}


\affiliation[inst4]{organization={School of Literature, Languages and Culture, University of Edinburgh},
            addressline={50 George Square}, 
            city={Edinburgh},
            postcode={EH89JY}, 
            country={United Kingdom}}  

\affiliation[inst5]{organization={Edinburgh Futures Institute, University of Edinburgh},
            addressline={1 Lauriston Place}, 
            city={Edinburgh},
            postcode={EH39EF}, 
            country={United Kingdom}} 

\affiliation[inst6]{organization={UK Dementia Research Institute, Centre at The University of Edinburgh},
            addressline={49 Little France Crescent}, 
            city={Edinburgh},
            postcode={EH164SB}, 
            country={United Kingdom}}

\affiliation[inst7]{organization={Usher Institute, University of Edinburgh},
            addressline={NINE, 9 Little France Road}, 
            city={Edinburgh},
            postcode={EH164UX}, 
            country={United Kingdom}}

\begin{abstract}

Cerebrovascular diseases (CVD) can lead to stroke and dementia. Stroke is the second leading cause of death world wide and dementia incidence is increasing by the year. There are several markers of CVD that are visible on brain imaging, including: white matter hyperintensities (WMH), acute and chronic ischaemic stroke lesions (ISL), lacunes, enlarged perivascular spaces (PVS), acute and chronic haemorrhagic lesions, and cerebral microbleeds (CMB). Brain atrophy also occurs in CVD. These markers are important for patient management and intervention, since they indicate elevated risk of future stroke and dementia.
We systematically reviewed automated systems designed to support radiologists reporting on these CVD imaging findings. We considered commercially available software and research publications which identify \textit{at least two} CVD markers.
In total, we included 29 commercial products and 13 research publications. Two distinct types of commercial support system were available: those which identify acute stroke lesions (haemorrhagic and ischaemic) from computed tomography (CT) scans, mainly for the purpose of patient triage; and those which measure WMH and atrophy regionally and longitudinally. In research, WMH and ISL were the markers most frequently analysed together, from magnetic resonance imaging (MRI) scans; lacunes and PVS were each targeted only twice and CMB only once.
For stroke, commercially available systems largely support the emergency setting, whilst research systems consider also follow-up and routine scans. The systems to quantify WMH and atrophy are focused on neurodegenerative disease support, where these CVD markers are also of significance. There are currently no openly validated systems, commercially, or in research, performing a comprehensive joint analysis of all CVD markers (WMH, ISL, lacunes, PVS, haemorrhagic lesions, CMB, and atrophy).

\end{abstract}

\begin{keyword}
Cerebrovascular disease \sep neuroradiology \sep machine learning
\end{keyword}

\end{frontmatter}

\section{Introduction}

Stroke is the second leading cause of death globally \citep{feigin_global_2021}, with 6.5 million deaths resulting from 12.2 million annual reported occurrences \citep{feigin_world_2022}. Given the high morbidity of stroke, with almost 50\% of sufferers being left disabled, prevention is of paramount importance. Epidemiological studies have identified several risk factors for stroke, most of which are modifiable. Amongst them are hypertension, diabetes mellitus, hypercholesterolaemia, cardiovascular diseases, and smoking \citep{donkor_stroke_2018}, all which have been associated with cerebrovascular disease (CVD) markers identifiable through different medical imaging modalities. CVD markers include ischaemic stroke lesions (ISL), haemorrhagic stroke lesions, white matter hyperintensities (WMH), cerebral microbleeds (CMB), lacunes, enlarged perivascular spaces (PVS), brain aneurysms (BA), and brain atrophy. In turn, previous stroke or the presence of CVD markers have been linked with increased risk of future stroke \citep{arboix_cardiovascular_2015,debette_clinical_2010}, while CVD markers have also been associated with dementia \citep{tanabe_white_2011}. CVD is thought to be a ``whole-brain disease'' \citep{shi_update_2016, duering2023neuroimaging}, therefore, the simultaneous identification of all possible CVD markers can play a valuable role in enabling appropriate preventative measures and lifestyle modifications.

\vspace{10pt}
There is a worldwide shortage of radiologists \citep{RSNA}, and software systems to support image interpretation are a potential solution. Not only this, but automation has the means to reduce the variability of interpretations seen amongst radiologists. CVD marker assessment differs by imaging modality, since different CVD marker subsets are visible in different imaging modalities. For instance, computed tomography (CT) scans may be used to assess lacunes and WMH, whilst magnetic resonance imaging (MRI) scans show more soft tissue detail and may be used to assess lacunes, WMH, PVS, CMB, and ISL. Any neuroradiological support system for CVD should provide information on all visible markers of CVD in the target image modality. In this systematic review, we define a \textit{complete CVD support system} as one which is able to identify and assess lacunes, WMH, PVS, CMB, and ISL from one or more medical imaging scans.

\vspace{10pt}
Previous reviews have examined computational neuroradiology support systems \citep{olthof_promises_2020, yearley_fda-approved_2023, yao_deep_2020}, stroke support systems \citep{wardlaw_accuracy_2022, soun_artificial_2021, bivard_artificial_2020, yeo_review_2021, mikhail_computational_2020, murray_artificial_2020,abbasi2023automatic}, WMH detection systems \citep{balakrishnan2021automatic}, CMB detection systems \citep{ferlin2023exploring}, and PVS detection systems \citep{barisano2022imaging,pham2022critical,waymont2024systematic}. \citet{jiang2022computer} performed a systematic review on computational extraction of three cerebral small vessel disease (cSVD) markers visible on MRI: PVS, CMB and lacunes. They found that good performance was achieved in small private datasets, but no pipeline has been validated in larger more heterogeneous datasets. 

Importantly, images from cSVD patients are likely to have multiple markers which confound each other due to similar appearance (e.g. PVS and lacunes or ISL and WMH), presenting a real challenge to automated detection. In this systematic review, we identify both commercially available systems and research publications on fully automatic methods, software products or platforms, which assess \emph{multiple} CVD markers, and meta-analyse their scope and characteristics.

\section{Methods}

We jointly perform a systematic review of commercial systems and research publications. In this section we describe our selection, search, and analysis methods.

\subsection{Selection strategy}

We selected relevant works according to the criteria below.

\paragraph{Inclusion criteria} The primary inclusion criterion was that the commercial system or primary publication must present a method, software or platform that \emph{fully automatically} assesses \emph{multiple} CVD markers from radiological brain images. Methods that only aimed at segmenting, quantifying or scoring 1 CVD feature, but in the process also identified other confounding CVD features for exclusion, were considered to qualify as identifying multiple CVD markers. This restriction on the number of CVD markers served to reduce the results to a manageable volume whilst focusing in on the most relevant systems for assessing such a complex, multifaceted disease. Research publications included only peer-reviewed full papers.

\paragraph{Exclusion criteria} The following were excluded: 

\begin{itemize}\setlength{\parskip}{0pt}
    \item Semi-automated methods
    \item Abstracts, where an associated full paper was not available (research systems)
    \item Non-peer-reviewed publications (research systems)
    \item Publications with no evaluation against ground truth (research systems)
    \item Non-English language publications (research systems)
    \item Products from companies with non-English language websites/materials without a translation available (commercial systems)
    \item Animal studies
\end{itemize}

\subsection{Search strategy}

We followed different strategies for identifying commercial systems and research publications, as detailed below.

\subsubsection{Searching for commercial systems}

We used multiple sources to review companies who were active in the period October 2022 to October 2024, which might develop products fitting the inclusion criteria. We reviewed companies:

\begin{itemize} \setlength{\parskip}{0pt}
    \item mentioned in \cite{wardlaw_accuracy_2022}'s evaluation of the accuracy of automated stroke diagnosis systems
    \item mentioned in \cite{olthof_promises_2020}'s review of AI in neuroradiology
    \item listed on the website \url{https://grand-challenge.org/aiforradiology/} (describing products related to AI in radiology) which were categorised as ``neuro''
    \item who had exhibited at the Radiological Society of North America (RSNA) (\url{https://www.rsna.org/}) in 2021, 2022 or 2023 (other conferences were also considered but rejected due to significant overlap) 
\end{itemize}

To decide whether any product from a company met the inclusion criteria, we carefully checked the websites and online materials related to the relevant product, as well as checking the US Food and Drug Administration (FDA) 510(k) pre-market notification documents where possible, since this often contains details not found elsewhere.

\subsubsection{Searching for research publications}

The literature search was carried out on Web of Science (\url{https://www.webofscience.com/wos/woscc/}). The search (conducted on 18/09/2024) included papers published between 2008 and September 2024, which mention at least 2 CVD features as well as referencing automation or machine learning. The start date was chosen because we are interested in systems with reasonable performance by today's standards, and 16 years of advancement in computer vision and the advent of deep learning likely render prior works - although potentially methodologically interesting - less relevant in the absence of a more recent update or enhancement.

The combination of search terms used in the advanced search was:

\vspace{10pt}
\noindent\emph{((white matter hyper* AND lacune*) OR (white matter hyper* AND (infarct OR stroke)) OR (white matter hyper* AND (perivascular spaces OR Virchow-Robin)) OR (white matter hyper* AND microbleed) OR (lacune* AND (infarct OR stroke)) OR (lacune* AND (perivascular spaces OR Virchow-Robin)) OR (lacune* AND microbleed) OR ((infarct OR stroke) AND (perivascular spaces OR Virchow-Robin)) OR ((infarct OR stroke) AND microbleed) OR ((perivascular spaces OR Virchow-Robin) AND microbleed)) AND (segment* OR classifier OR machine learning OR deep learning OR ai OR auto* quantif*)}
\vspace{10pt}

Abstracts were screened to identify articles likely to meet the inclusion criteria. The full text of these articles was then inspected to produce the final list of included research publications. Articles from authors' personal libraries known to meet the inclusion criteria and not identified in the literature search were also added to the final list.

\subsection{Data extraction}

For each neuroradiology support system, we extracted information regarding: the modalities/sequences used; the pathologies targeted; the specific features of the system (e.g.~segmentation); development and validation data characteristics and reference standard; the method/training details; the validation details; and obtained regulatory approvals (for commercial systems only).

For the commercial systems, the information comes from the company website, white papers, explicitly cited publications, and FDA pre-market notifications. Where information was lacking about methods and training data, a further search was performed including review of linked publications. Where a company offered sub-products making up a more comprehensive product, the complete product was considered, and the features were considered to be the combination of the features of the sub-products. However, where sensible, information was reported for each sub-product.

\subsection{Analysis of bias risks and applicability}
We used a version of the QUADAS 2 tool which is used to assess risk of bias and applicability in systematic reviews.\footnote{\url{https://www.bristol.ac.uk/media-library/sites/quadas/migrated/documents/quadas2.pdf}} This table is commonly presented with the entries ``low", ``unclear", and ``high". However, since a high risk of bias is bad, but a high applicability is good, we chose to use ``good'', ``medium'', and ``bad'' as categories. An example of an entry where ``bad'' would be given is in the question ``Is the ref std. likely to be correct?'' if the ground truth was automatically generated.

\section{Results}

\subsection{Search results}

\begin{figure}[!htb]
    \vspace{10pt}
    \begin{subfigure}{\textwidth}
        \centering
        \includegraphics[width=\textwidth]{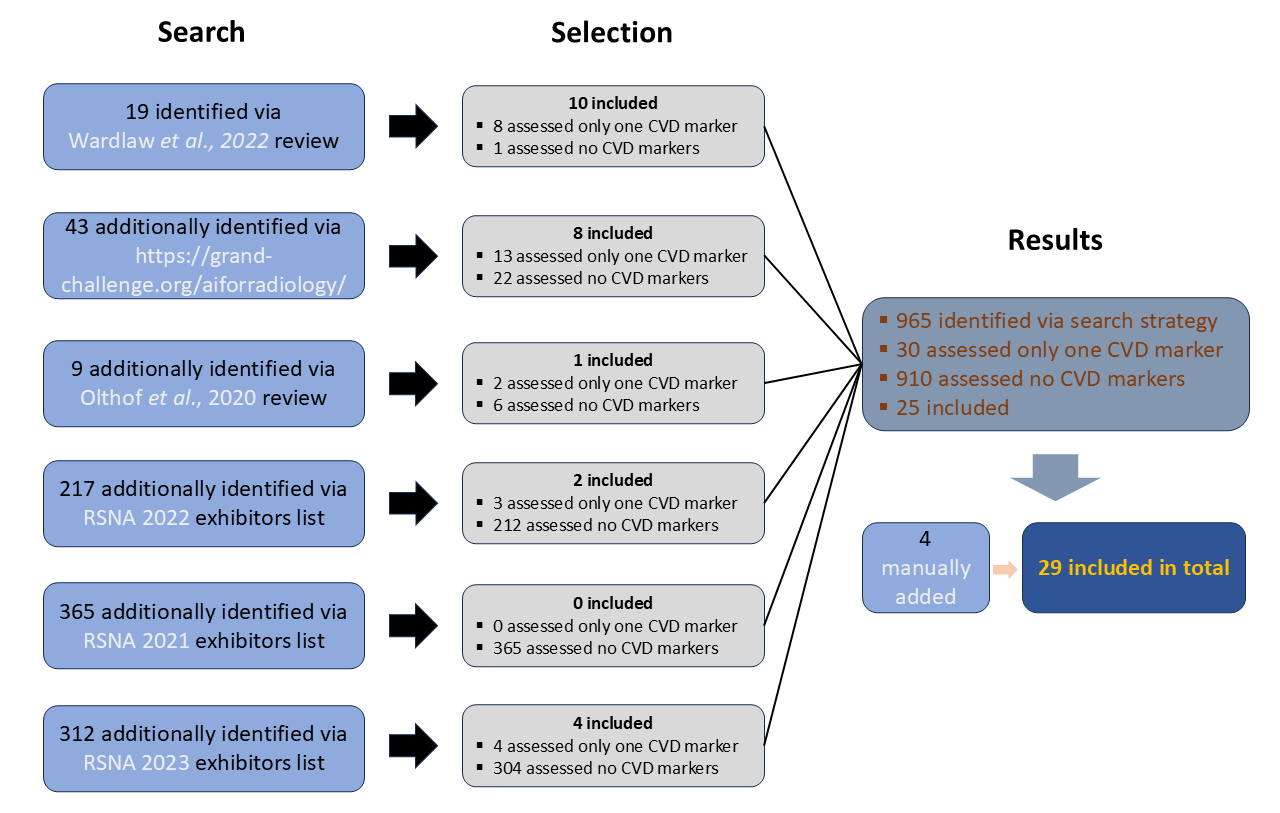}
        \caption{Data collection process and results for commercially available CVD support systems with selection based on the inclusion/exclusion criteria}
    \end{subfigure}
    \hfill
    \begin{subfigure}{\textwidth}
        \centering
        \includegraphics[width=\textwidth]{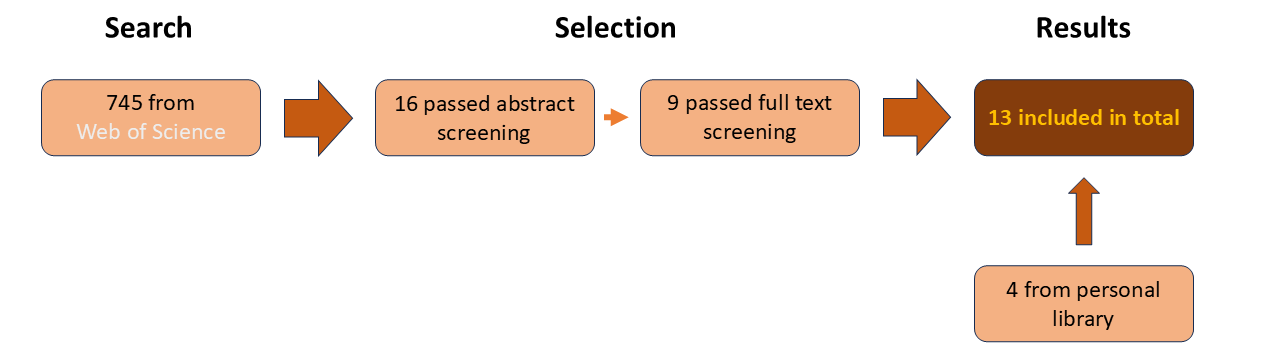}
        \caption{Data collection process and results for research publications on CVD support systems with selection based on the inclusion/exclusion criteria}
    \end{subfigure}
    \caption{}
    \label{fig:collection_flow}
\end{figure}

\textbf{Fig \ref{fig:collection_flow}} shows the results of the data collection process. In total, 965 commercial systems were identified as candidates, but after applying the inclusion criteria, only 25 remained. 30 of the 965 systems were rejected because they evaluated only 1 CVD marker and the remainder were rejected because they did not evaluate any of the CVD markers. In total 29 systems were considered after manually including 4 systems known to the authors, which were not identified as part of the search strategy. After screening the abstracts of 745 research publications, 16 were selected for full text screening and 9 passed the inclusion criteria. Additionally, 4 papers from authors' personal libraries were included.

\subsection{Commercial systems}

The results of reviewing the 29 identified products can be seen in \textbf{Table \ref{tab:commercial1}}. Products broadly fall into 2 categories: stroke detection systems and general neuroradiology support systems with emphasis on neurodegenerative diseases. Below we summarise the types of CVD markers and imaging modalities that these products support, characteristics of the data used in development/testing, deep learning methods applied, regulatory approval, and the availability of validation information. More detail, including information about the reference standard and product impact, is provided in the supplementary material.

\subsubsection{CVD markers and imaging modalities}

\begin{figure}[!htb]
    \begin{subfigure}{0.49\textwidth}
        \centering
        \includegraphics[width=\textwidth]{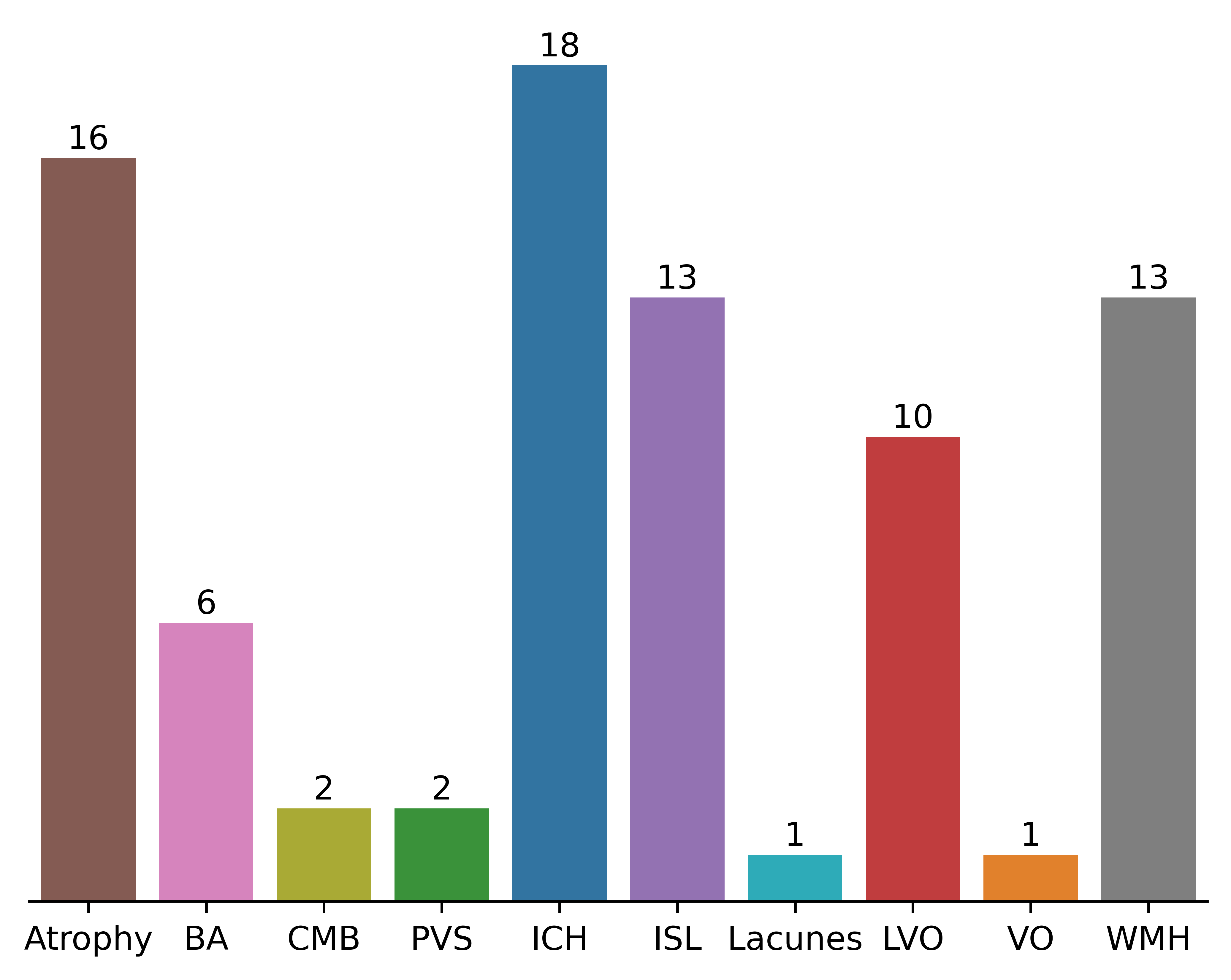}
        \caption{Number of systems targeting each pathology.}
        \label{fig:path dist commercial bar}
    \end{subfigure}
    \hfill
    \begin{subfigure}{0.49\textwidth}
        \centering
        \includegraphics[width=\textwidth]{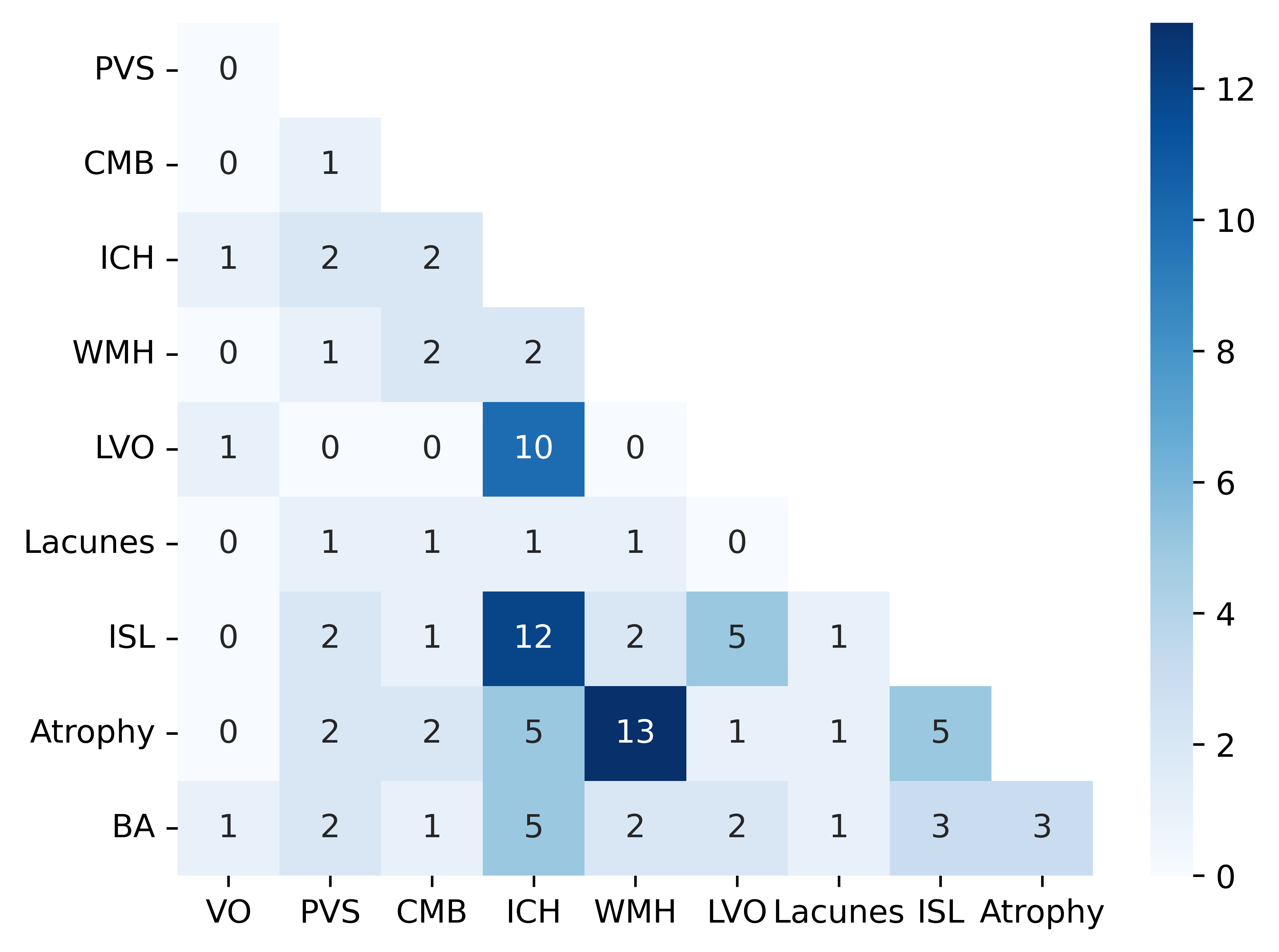}
        \caption{Co-occurrence of pathologies targeted.}
        \label{fig:path dist commercial matrix}
    \end{subfigure}
    \caption{Distribution of pathologies targeted by the identified commercial systems.}
    \label{fig:path dist commercial}
\end{figure}

\paragraph{Stroke systems} The 19 stroke detection systems (Always-on AI, Enterprise CTB, Cina, e-Stroke, BrainScan CT, Cercare Stroke, APOLLO BRAIN, DeepCT, icobrain, NeuroShield CT-TBI, MEDIHUB STROKE, Stroke Suite, Stroke-Viewer, qER, Rapid Stroke, Digital Brain, Viz Radiology Suite, StroCare Suite, uAI Discover IschaemicStroke/ICH) place emphasis on rapid identification and triage of acute stroke, i.e.~identifying ischaemic stroke lesions (ISL) and intracranial haemorrhages (ICH). This can be seen in the co-occurrence matrix in \textbf{Fig \ref{fig:path dist commercial matrix}} displaying the number of paired CVD markers, which shows that 12 systems assess both haemorrhagic strokes and ischaemic stroke lesions (i.e.~ICH and ISL), and 10 assess both haemorrhagic strokes and occlusion in large vessels (i.e.~ICH and LVO). Both ISL and large vessel occlusion (LVO) are indicators of ischaemic stroke; LVO detection identifies the location of the blocked vessel and specifically the blockage (usually from CT angiogram (CTA) which uses contrast injection to show the vessels), whilst ISL detection systems show areas of ischaemic tissue (which may not necessarily be limited to the site of the occlusion).

Consistent with the interest in diagnosing the cause of acute stroke, 17/19 systems target non-contrast CT (NCCT), which is a widely used imaging modality for acute stroke world wide. Of these, 8/17 systems assess the Alberta stroke program early CT score (ASPECTS) \citep{barber2000validity} and outline the area of hypoattenuation (which corresponds to the ISL marker). Then, 8/17 systems target CT perfusion (CTP) to segment the ISL core. Finally, 10 systems target CT angiograms (CTA) for detecting or localising occlusion in large vessels (LVO), and 4 systems detect aneurysms from CTA. 

Overall, Enterprise CTB from annalise.ai appears to be the most comprehensive system, detecting 6 CVD markers amongst the 130 it assesses in NCCT scans. However, this includes PVS which is typically best identified on MRI (not NCCT), and there is no available validation data on PVS and some of the other markers. uAI Discover from United Imaging Intelligence also claims to assess 6 CVD markers, although no further information regarding the product's application or validation was found.

\paragraph{General neuroradiology systems} There were 16 systems that quantify brain atrophy and/or WMH (Enterprise CTB, BrainScan CT, cMRI, NeuroQuant, icobrain, NeuroShield MR, mdbrain, AQUA, Pixyl.Neuro, Quantib ND, NEUROCLOUD VOL, DeepBrain, AgingCare Suite, QP-Brain, QyScore, UAI Discover CSVD). Although these systems are mostly intended to support diagnosis of diseases such as dementia, they have been included in our review since brain atrophy and WMH are also markers of CVD. All these systems except for Enterprise CTB and BrainScan CT use MRI (T1 and/or FLAIR explicitly mentioned for all products with the exception of NeuroShield, QUANTIB ND and UAI Discover CSVD). 12/16 systems provide atrophy quantification by region (9 perform longitudinal analysis) and the same number provide WMH segmentation and volumes. 5 systems provide regional FLAIR-hyperintense volumes in periventricular, juxtacortical/cortical, infratentorial, and the spinal cord (if any).

\subsubsection{Sample characteristics}

\paragraph{Development}
Information on training/development data was found for at least 1 sub-product in 15/29 systems. The median dataset size was 2869 subjects and the maximum was 12.5 million (BrainScan CT). In 9/15 systems, the use of data from multiple scanners or sites was stated for at least 1 sub-product, while at least 1 sub-product in 2/15 and 3/15 described the age and sex characteristics respectively.

\paragraph{Validation} Information on validation data was found for at least 1 sub-product in 19/29 systems. The median dataset size was 270 subjects and the maximum was 3363 (Enterprise CTB). The use of multiple scanners or sites was mentioned by at least 1 sub-product in 12/19 systems and the sex and age characteristics were detailed for at least 1 sub-product by 7/19.

\subsubsection{Deep learning methods}

18/29 systems utilise deep learning (DL) methods in at least 1 of their sub-products. Of these, all deploy convolutional neural networks (CNNs) \citep{lecun_backpropagation_1989}, with 1 product (Enterprise CTB) also making use of the vision transformer architecture (VIT) \citep{dosovitskiy_image_2021}. Of the 27 sub-products using CNNs, 9 state the use of a U-Net based model \citep{ronneberger_u-net_2015} and 9 state that a 3D architecture is used. 9 sub-products utilised multiple models with at least 1 being a DL model, in multi-stage pipelines or ensembles. Of these, 7 described the purpose of each model. Details regarding at least 1 aspect of the training process (e.g.~data augmentation, loss function, optimiser, data sampling, etc) were found for 5 sub-products. Of the 19 systems assessing stroke, 15 utilise DL in at least 1 sub-product while 5/16 of the systems offering WMH or atrophy segmentation/quantification use DL.


\subsubsection{Validation and regulatory approval}

A full analysis of all internal or external validation studies for each product was outside the scope of this review, but we refer the interested reader to \citet{yearley_fda-approved_2023}.

Of the 50 sub-products for which validation metrics were found, 28 provided sensitivity and specificity (13 with confidence intervals), 9 reported area under the receiver operating characteristic curve (AUC), and 17 reported Dice similarity coefficient (DSC). The most frequently claimed clinical impact was a reduction in time required for the patient workflow, which was stated by 12 subsystems.

Of the 29 systems identified, 19 have at least 1 sub-product which is FDA approved and 26 have at least 1 sub-product with CE marking.

\tiny
\begin{longtable}[c]{lllllll}
\caption{Commercial systems - overview. Some products include additional non-CVD related features not mentioned here (e.g. assessment of cranial fracture or glioma).}
\label{tab:commercial1}\\
\multicolumn{7}{l}{\begin{tabular}[c]{@{}l@{}}Pathologies and anatomy:\\ ICH: intracerebral haemorrhage, (L)VO: (large) vessel occlusion, BA: brain aneurysm, \\ ISL: ischaemic stroke lesion, PVS: enlarged perivascular spaces, SVD: small vessel disease,\\ WMH: white matter hyperintensities, EDH: epidural hematoma, SDH: subdural haemorrhage,\\ SAH: subarachnoid haemorrhage, IVH: intraventricular haemorrhage, IPH: intraparenchymal \\ haemorrhage, ICA: internal carotid artery, MCA: middle cerebral artery.\\ \\ Modalities:\\ NCCT: non-contrast computed tomography, CTA: computed tomography angiography, \\ CTP: computed tomography perfusion, MRI: magnetic resonance imaging, T1: T1-weighted\\ MRI, T2: T2-weighted MRI, T2*: T2*-weighted MRI, FLAIR: fluid-attenuated inversion\\ recovery MRI, SWI: susceptibility-weighted imaging MRI, DWI: diffusion-weighted imaging\\ MRI, GRE: gradient echo MRI, TOF: time-of-flight MRI angiography.\\ \\ Other abbreviations:\\ AI: artificial intelligence, DL: deep learning, ML: machine learning, RB: rule-based.\\ \\ (AI is stated as the method class when the company states the product as being AI and further \\ investigation does not clarify the method).\end{tabular}} \\ \hline
\endfirsthead
\multicolumn{7}{c}%
{{\bfseries Table \thetable\ continued from previous page}} \\
\endhead
\multicolumn{1}{|l|}{\textbf{Company}} &
  \multicolumn{1}{l|}{\textbf{Product}} &
  \multicolumn{1}{l|}{\textbf{Pathology}} &
  \multicolumn{1}{l|}{\textbf{Modality}} &
  \multicolumn{1}{l|}{\textbf{Features}} &
  \multicolumn{1}{l|}{\textbf{\begin{tabular}[c]{@{}l@{}}Method\\ class\end{tabular}}} &
  \multicolumn{1}{l|}{\textbf{\begin{tabular}[c]{@{}l@{}}Approval\\ (sub-product)\end{tabular}}} \\ \hline
\multicolumn{1}{|l|}{\begin{tabular}[c]{@{}l@{}}aidoc, \\ icometrix\end{tabular}} &
  \multicolumn{1}{l|}{\begin{tabular}[c]{@{}l@{}}Always-on \\ AI\end{tabular}} &
  \multicolumn{1}{l|}{\begin{tabular}[c]{@{}l@{}}ICH, VO, \\ M1 LVO, \\ BA\end{tabular}} &
  \multicolumn{1}{l|}{\begin{tabular}[c]{@{}l@{}}NCCT,\\ CTA,\\ CTP\end{tabular}} &
  \multicolumn{1}{l|}{\begin{tabular}[c]{@{}l@{}}Identification of \\ suspected findings\\ per pathology. \\ Core segmentation \\ and volume for \\ (M1 L)VO from \\ CTP.\end{tabular}} &
  \multicolumn{1}{l|}{DL} &
  \multicolumn{1}{l|}{\begin{tabular}[c]{@{}l@{}}FDA (II), CE (I)\\ (BriefCase ICI \\ triage)\\ (BriefCase LVO \\ triage)\\ (Briefcase BA \\ triage) \\ (icobrain-cva)\end{tabular}} \\ \hline
\multicolumn{1}{|l|}{annalise.ai} &
  \multicolumn{1}{l|}{\begin{tabular}[c]{@{}l@{}}Enterprise \\ CTB\end{tabular}} &
  \multicolumn{1}{l|}{\begin{tabular}[c]{@{}l@{}}ICH, ISL,\\ BA,\\ atrophy,\\ PVS, \\ SVD,\\ ...\\ (130 total)\end{tabular}} &
  \multicolumn{1}{l|}{NCCT} &
  \multicolumn{1}{l|}{\begin{tabular}[c]{@{}l@{}}Detection or seg-\\ mentation of 130 \\ radiological \\ findings including \\ ICH subtypes, \\ infarcts, PVS, and \\ SVD.\end{tabular}} &
  \multicolumn{1}{l|}{DL} &
  \multicolumn{1}{l|}{FDA (II), CE (IIb)} \\ \hline
\multicolumn{1}{|l|}{Avicenna.AI} &
  \multicolumn{1}{l|}{Cina} &
  \multicolumn{1}{l|}{\begin{tabular}[c]{@{}l@{}}ICH, LVO,\\ ISL\end{tabular}} &
  \multicolumn{1}{l|}{\begin{tabular}[c]{@{}l@{}}NCCT,\\ CTA\end{tabular}} &
  \multicolumn{1}{l|}{\begin{tabular}[c]{@{}l@{}}Identification of \\ suspected ICH or \\ LVO. ASPECTS \\ with infarct outline\\ and probability map \\ of hypoattenuation \\ and sulcal efface-\\ ment.\end{tabular}} &
  \multicolumn{1}{l|}{DL} &
  \multicolumn{1}{l|}{\begin{tabular}[c]{@{}l@{}}FDA (II), CE (I)\\ (CINA-ICH)\\ (CINA-LVO)\\ \\ CE (I)\\ (CINA-\\ ASPECTS)\end{tabular}} \\ \hline
\multicolumn{1}{|l|}{BRAINOMIX} &
  \multicolumn{1}{l|}{e-Stroke} &
  \multicolumn{1}{l|}{\begin{tabular}[c]{@{}l@{}}ICH, LVO,\\ ISL\end{tabular}} &
  \multicolumn{1}{l|}{\begin{tabular}[c]{@{}l@{}}NCCT,\\ CTA,\\ CTP\end{tabular}} &
  \multicolumn{1}{l|}{\begin{tabular}[c]{@{}l@{}}ASPECTS with \\ infarct volume \\ and hypoattenuation \\ heat map (hyper-\\ attenuation also \\ detected indicating \\ bleeding). \\ Bounding box \\ on CTA for vessel \\ occlusion. Core \\ and penumbra \\ segmentation and \\ volume and mis-\\ match ratio from \\ CTP.\end{tabular}} &
  \multicolumn{1}{l|}{\begin{tabular}[c]{@{}l@{}}ML,\\ RB,\\ DL\end{tabular}} &
  \multicolumn{1}{l|}{\begin{tabular}[c]{@{}l@{}}FDA (II), CE (IIa)\\ (e-CTA)\\ \\ CE (IIa)\\ (e-CTP)\\ (e-ASPECTS)\end{tabular}} \\ \hline
\multicolumn{1}{|l|}{BRAINSCAN.AI} &
  \multicolumn{1}{l|}{\begin{tabular}[c]{@{}l@{}}BrainScan \\ CT\end{tabular}} &
  \multicolumn{1}{l|}{\begin{tabular}[c]{@{}l@{}}ICH, ISL,\\ atrophy\end{tabular}} &
  \multicolumn{1}{l|}{NCCT} &
  \multicolumn{1}{l|}{\begin{tabular}[c]{@{}l@{}}Segmentation and \\ probability map of \\ ICH subtypes and \\ acute and chronic \\ ISL as well as areas\\ of atrophy.\end{tabular}} &
  \multicolumn{1}{l|}{AI} &
  \multicolumn{1}{l|}{CE (IIa)} \\ \hline
\multicolumn{1}{|l|}{Cercare Medical} &
  \multicolumn{1}{l|}{\begin{tabular}[c]{@{}l@{}}Cercare \\ Stroke\end{tabular}} &
  \multicolumn{1}{l|}{ICH, LVO} &
  \multicolumn{1}{l|}{\begin{tabular}[c]{@{}l@{}}NCCT,\\ CTA,\\ MRI\end{tabular}} &
  \multicolumn{1}{l|}{\begin{tabular}[c]{@{}l@{}}Detects ICH. \\ Identifies \\ and provides \\ volume of \\ infarcted lesions. \\ Identifies exact \\ location of LVO. \\ Perfusion maps \\ from CT or MRI \\ perfusion \\ imaging and \\ additional \\ biomarkers of \\ oxygenation.\end{tabular}} &
  \multicolumn{1}{l|}{DL} &
  \multicolumn{1}{l|}{CE (IIa)} \\ \hline
\multicolumn{1}{|l|}{CEREBRIU} &
  \multicolumn{1}{l|}{\begin{tabular}[c]{@{}l@{}}APOLLO \\ BRAIN\end{tabular}} &
  \multicolumn{1}{l|}{ICH, ISL} &
  \multicolumn{1}{l|}{\begin{tabular}[c]{@{}l@{}}FLAIR,\\ T2*,\\ SWI,\\ DWI\end{tabular}} &
  \multicolumn{1}{l|}{\begin{tabular}[c]{@{}l@{}}Detects ICH and \\ ISL.\end{tabular}} &
  \multicolumn{1}{l|}{AI} &
  \multicolumn{1}{l|}{CE (I)} \\ \hline
\multicolumn{1}{|l|}{Combinostics} &
  \multicolumn{1}{l|}{cMRI} &
  \multicolumn{1}{l|}{\begin{tabular}[c]{@{}l@{}}WMH,\\ Atrophy\end{tabular}} &
  \multicolumn{1}{l|}{\begin{tabular}[c]{@{}l@{}}T1, \\ FLAIR\end{tabular}} &
  \multicolumn{1}{l|}{\begin{tabular}[c]{@{}l@{}}Segmentation of\\ brain and atrophy \\ per region and \\ WMH per region; \\ longitudinal analysis\\ available. Outputs\\ structured report.\end{tabular}} &
  \multicolumn{1}{l|}{RB} &
  \multicolumn{1}{l|}{FDA (II), CE (IIa)} \\ \hline
\multicolumn{1}{|l|}{cortechs.ai} &
  \multicolumn{1}{l|}{\begin{tabular}[c]{@{}l@{}}Neuro-\\ Quant\end{tabular}} &
  \multicolumn{1}{l|}{\begin{tabular}[c]{@{}l@{}}WMH,\\ atrophy\end{tabular}} &
  \multicolumn{1}{l|}{\begin{tabular}[c]{@{}l@{}}T1,\\ FLAIR\end{tabular}} &
  \multicolumn{1}{l|}{\begin{tabular}[c]{@{}l@{}}Brain atrophy per \\ region from T1. \\ Segmentation and \\ volume of WMH \\ and longitudinal \\ analysis from \\ FLAIR.\end{tabular}} &
  \multicolumn{1}{l|}{\begin{tabular}[c]{@{}l@{}}ML,\\ RB,\\ AI\end{tabular}} &
  \multicolumn{1}{l|}{FDA (II), CE (IIa)} \\ \hline
\multicolumn{1}{|l|}{DEEP01} &
  \multicolumn{1}{l|}{DeepCT} &
  \multicolumn{1}{l|}{ICH, LVO} &
  \multicolumn{1}{l|}{\begin{tabular}[c]{@{}l@{}}NCCT,\\ CTA\end{tabular}} &
  \multicolumn{1}{l|}{\begin{tabular}[c]{@{}l@{}}Detection of ICH \\ (region, subtype, \\ volume, modified\\ Fisher scale, \\ Marshall score,\\ IMPACT score, \\ midline shift, and \\ Evans index).\\ Detection and \\ localisation of \\ LVO.\end{tabular}} &
  \multicolumn{1}{l|}{DL} &
  \multicolumn{1}{l|}{FDA (II), CE (I)} \\ \hline
\multicolumn{1}{|l|}{icometrix} &
  \multicolumn{1}{l|}{icobrain} &
  \multicolumn{1}{l|}{\begin{tabular}[c]{@{}l@{}}ISL,\\ WMH, \\ atrophy\end{tabular}} &
  \multicolumn{1}{l|}{\begin{tabular}[c]{@{}l@{}}CTP,\\ T1,\\ FLAIR\end{tabular}} &
  \multicolumn{1}{l|}{\begin{tabular}[c]{@{}l@{}}Segmentation of \\ ischaemic core and \\ penumbra from \\ CTP. Segmen-\\ tation and volume \\ and longitudinal \\ analysis of FLAIR \\ hyperintensities and\\ T1 hypointensities.  \\ Brain volume \\ change metrics by \\ region from T1.\end{tabular}} &
  \multicolumn{1}{l|}{AI} &
  \multicolumn{1}{l|}{FDA (II), CE (I)} \\ \hline
\multicolumn{1}{|l|}{INMED AI} &
  \multicolumn{1}{l|}{NeuroShield} &
  \multicolumn{1}{l|}{\begin{tabular}[c]{@{}l@{}}ICH,\\ WMH, \\ CMB,\\ atrophy\end{tabular}} &
  \multicolumn{1}{l|}{\begin{tabular}[c]{@{}l@{}}NCCT,\\ MRI\end{tabular}} &
  \multicolumn{1}{l|}{\begin{tabular}[c]{@{}l@{}}Brain atrophy by \\ region, WMH and\\ CMBs (no info),\\ detection and seg-\\ mentation of ICH\\ and triage.\end{tabular}} &
  \multicolumn{1}{l|}{DL} &
  \multicolumn{1}{l|}{\begin{tabular}[c]{@{}l@{}}FDA (II)\\ (NeuroShieldMR)\end{tabular}} \\ \hline
\multicolumn{1}{|l|}{JLK} &
  \multicolumn{1}{l|}{\begin{tabular}[c]{@{}l@{}}MEDIHUB \\ STROKE\end{tabular}} &
  \multicolumn{1}{l|}{ICH, ISL} &
  \multicolumn{1}{l|}{\begin{tabular}[c]{@{}l@{}}NCCT,\\ CTP, T1,\\ T2,\\ FLAIR,\\ GRE,\\ PWI,\\ DWI\end{tabular}} &
  \multicolumn{1}{l|}{\begin{tabular}[c]{@{}l@{}}ICH localisation \\ and subtype (EDH,\\ SDH, SAH, IVH, \\ intracerebral) from\\ NCCT. Ischaemic \\ stroke subtype \\ probability  \\ from DWI and \\ optional FLAIR, T1,\\ T2. Severity \\ probability \\ according\\ to the NIH stroke \\ scale from DWI. \\ Volume mismatch \\ from DWI/PWI. \\ On-set time \\ prediction and \\ ischaemic core seg-\\ mentation from \\ DWI/FLAIR. \\ Collateral map \\ analysis, ischaemic \\ core and penumbra \\ volume and mis-\\ match from CTP. \\ Infarct segmentation \\ and prognosis from\\ DWI and early \\ detection from \\ NCCT.\end{tabular}} &
  \multicolumn{1}{l|}{\begin{tabular}[c]{@{}l@{}}DL,\\ AI\end{tabular}} &
  \multicolumn{1}{l|}{\begin{tabular}[c]{@{}l@{}}CE (I)\\ (JBS-01K)\\ (JBS-04K)\end{tabular}} \\ \hline
\multicolumn{1}{|l|}{mediaire} &
  \multicolumn{1}{l|}{mdbrain} &
  \multicolumn{1}{l|}{\begin{tabular}[c]{@{}l@{}}WMH,\\ atrophy,\\ BA\end{tabular}} &
  \multicolumn{1}{l|}{\begin{tabular}[c]{@{}l@{}}T1,\\ FLAIR,\\ TOF\end{tabular}} &
  \multicolumn{1}{l|}{\begin{tabular}[c]{@{}l@{}}Brain volume by \\ region and long-\\ itudinal changes \\ from T1. Segmen-\\ tation of WMH \\ and regional \\ assessment and \\ longitudinal\\ changes from T1\\ /FLAIR. Segmen-\\ tation of BA from \\ TOF.\end{tabular}} &
  \multicolumn{1}{l|}{DL} &
  \multicolumn{1}{l|}{CE (I)} \\ \hline
\multicolumn{1}{|l|}{Methinks} &
  \multicolumn{1}{l|}{\begin{tabular}[c]{@{}l@{}}Stroke \\ Suite\end{tabular}} &
  \multicolumn{1}{l|}{ICH, LVO} &
  \multicolumn{1}{l|}{\begin{tabular}[c]{@{}l@{}}NCCT,\\ CTA\end{tabular}} &
  \multicolumn{1}{l|}{\begin{tabular}[c]{@{}l@{}}Notification of ICH \\ and LVO for \\ patient triage.\end{tabular}} &
  \multicolumn{1}{l|}{DL} &
  \multicolumn{1}{l|}{CE (I)} \\ \hline
\multicolumn{1}{|l|}{neurophet} &
  \multicolumn{1}{l|}{AQUA} &
  \multicolumn{1}{l|}{\begin{tabular}[c]{@{}l@{}}WMH,\\ atrophy\end{tabular}} &
  \multicolumn{1}{l|}{\begin{tabular}[c]{@{}l@{}}T1,\\ FLAIR\end{tabular}} &
  \multicolumn{1}{l|}{\begin{tabular}[c]{@{}l@{}}Brain atrophy by \\ region and longit-\\ udinal analysis \\ from T1. \\ WMH burden score\\ from FLAIR.\end{tabular}} &
  \multicolumn{1}{l|}{DL} &
  \multicolumn{1}{l|}{CE (IIa)} \\ \hline
\multicolumn{1}{|l|}{NICOLAB} &
  \multicolumn{1}{l|}{\begin{tabular}[c]{@{}l@{}}Stroke-\\ Viewer\end{tabular}} &
  \multicolumn{1}{l|}{\begin{tabular}[c]{@{}l@{}}ICH, LVO,\\ ISL\end{tabular}} &
  \multicolumn{1}{l|}{\begin{tabular}[c]{@{}l@{}}NCCT,\\ CTA,\\ CTP\end{tabular}} &
  \multicolumn{1}{l|}{\begin{tabular}[c]{@{}l@{}}Segmentation and\\ volume of  ICH \\ and SAH from \\ NCCT. ASPECTS \\ scoring. Identific-\\ ation and bounding\\ box of LVO and \\ collateral score \\ from CTA. Core \\ and mismatch \\ volumes from CTP.\end{tabular}} &
  \multicolumn{1}{l|}{\begin{tabular}[c]{@{}l@{}}DL,\\ AI\end{tabular}} &
  \multicolumn{1}{l|}{FDA (II), CE (I)} \\ \hline
\multicolumn{1}{|l|}{PIXYL} &
  \multicolumn{1}{l|}{Pixyl.Neuro} &
  \multicolumn{1}{l|}{\begin{tabular}[c]{@{}l@{}}WMH,\\ atrophy\end{tabular}} &
  \multicolumn{1}{l|}{\begin{tabular}[c]{@{}l@{}}T1, \\ FLAIR\end{tabular}} &
  \multicolumn{1}{l|}{\begin{tabular}[c]{@{}l@{}}Segmentation and\\ volume of WMH\\ by region\\ and longitudinally\\ and brain atrophy \\ by region and \\ longitudinally.\end{tabular}} &
  \multicolumn{1}{l|}{AI} &
  \multicolumn{1}{l|}{FDA (II), CE (IIa)} \\ \hline
\multicolumn{1}{|l|}{Quantib} &
  \multicolumn{1}{l|}{\begin{tabular}[c]{@{}l@{}}Quantib \\ ND\end{tabular}} &
  \multicolumn{1}{l|}{\begin{tabular}[c]{@{}l@{}}WMH,\\ atrophy\end{tabular}} &
  \multicolumn{1}{l|}{MRI} &
  \multicolumn{1}{l|}{\begin{tabular}[c]{@{}l@{}}Brain atrophy by \\ region and longit-\\ udinal changes. \\ Segmentation of \\ WMH and longit-\\ udinal changes.\end{tabular}} &
  \multicolumn{1}{l|}{ML} &
  \multicolumn{1}{l|}{FDA (II), CE (IIa)} \\ \hline
\multicolumn{1}{|l|}{QUBIOtech} &
  \multicolumn{1}{l|}{\begin{tabular}[c]{@{}l@{}}NEURO-\\ CLOUD \\ VOL\end{tabular}} &
  \multicolumn{1}{l|}{\begin{tabular}[c]{@{}l@{}}WMH,\\ atrophy\end{tabular}} &
  \multicolumn{1}{l|}{\begin{tabular}[c]{@{}l@{}}T1,\\ FLAIR\end{tabular}} &
  \multicolumn{1}{l|}{\begin{tabular}[c]{@{}l@{}}Brain atrophy \\ quantification by \\ region and longit-\\ udinal changes \\ from T1. Segmen-\\ tation of WMH by \\ region and longit-\\ udinal changes \\ from FLAIR.\end{tabular}} &
  \multicolumn{1}{l|}{\begin{tabular}[c]{@{}l@{}}ML,\\ RB\end{tabular}} &
  \multicolumn{1}{l|}{CE (I)} \\ \hline
\multicolumn{1}{|l|}{qure.ai} &
  \multicolumn{1}{l|}{qER} &
  \multicolumn{1}{l|}{ICH, ISL} &
  \multicolumn{1}{l|}{NCCT} &
  \multicolumn{1}{l|}{\begin{tabular}[c]{@{}l@{}}Bounding box and \\ volume for ICH \\ subtypes (IPH, IVH,\\  SDH, SAH, EDH). \\ Detects and quanti-\\ fies midline shift. \\ ASPECTS score \\ and outlines hypo-\\ attenuated area.\\ Bounding box for \\ acute and chronic \\ infarcts (including \\ lacunar).\end{tabular}} &
  \multicolumn{1}{l|}{DL} &
  \multicolumn{1}{l|}{FDA (II), CE (IIa)} \\ \hline
\multicolumn{1}{|l|}{RapidAI} &
  \multicolumn{1}{l|}{\begin{tabular}[c]{@{}l@{}}Rapid \\ Stroke\end{tabular}} &
  \multicolumn{1}{l|}{\begin{tabular}[c]{@{}l@{}}ICH, LVO,\\ ISL\end{tabular}} &
  \multicolumn{1}{l|}{\begin{tabular}[c]{@{}l@{}}NCCT,\\ CTA,\\ CTP\end{tabular}} &
  \multicolumn{1}{l|}{\begin{tabular}[c]{@{}l@{}}Identification of \\ ICH and segment-\\ ation and volume \\ of hyperattenuated \\ region from NCCT. \\ ASPECTS score. \\ Colour coded \\ overlay \\ of regions of blood \\ vessel density \\ asymmetry from \\ CTA. Identification\\ of LVO in the ICA \\ or MCA. Hypo-\\ attenuation and \\ mismatch volume \\ from CTP.\end{tabular}} &
  \multicolumn{1}{l|}{\begin{tabular}[c]{@{}l@{}}ML,\\ \\ DL,\\ AI\end{tabular}} &
  \multicolumn{1}{l|}{FDA (II), CE (I)} \\ \hline
\multicolumn{1}{|l|}{SHUKUN} &
  \multicolumn{1}{l|}{\begin{tabular}[c]{@{}l@{}}Digital \\ Brain\end{tabular}} &
  \multicolumn{1}{l|}{\begin{tabular}[c]{@{}l@{}}ICH, BA,\\ ISL\end{tabular}} &
  \multicolumn{1}{l|}{\begin{tabular}[c]{@{}l@{}}NCCT,\\ CTA\end{tabular}} &
  \multicolumn{1}{l|}{\begin{tabular}[c]{@{}l@{}}Segments ICH and \\ ischaemia from \\ NCCT. Provides \\ diameter measure-\\ ments and \\ ASPECTS. \\ Measures \\ dimensions of \\ BA from CTA.\end{tabular}} &
  \multicolumn{1}{l|}{AI} &
  \multicolumn{1}{l|}{CE (I)} \\ \hline
\multicolumn{1}{|l|}{Viz.ai} &
  \multicolumn{1}{l|}{\begin{tabular}[c]{@{}l@{}}Viz \\ Radiology \\ Suite\end{tabular}} &
  \multicolumn{1}{l|}{\begin{tabular}[c]{@{}l@{}}ICH, LVO,\\ BA\end{tabular}} &
  \multicolumn{1}{l|}{\begin{tabular}[c]{@{}l@{}}NCCT,\\ CTA\end{tabular}} &
  \multicolumn{1}{l|}{\begin{tabular}[c]{@{}l@{}}Detection of LVO \\ and BA from CTA. \\ Detection of ICH \\ and SDH from \\ NCCT.\end{tabular}} &
  \multicolumn{1}{l|}{DL} &
  \multicolumn{1}{l|}{\begin{tabular}[c]{@{}l@{}}FDA (II), CE (I)\\ (Viz LVO)\\ (Viz ICH)\\ \\ FDA (II)\\ (Viz \\ ANEURYSM)\end{tabular}} \\ \hline
\multicolumn{1}{|l|}{VUNO} &
  \multicolumn{1}{l|}{DeepBrain} &
  \multicolumn{1}{l|}{\begin{tabular}[c]{@{}l@{}}WMH,\\ Atrophy\end{tabular}} &
  \multicolumn{1}{l|}{\begin{tabular}[c]{@{}l@{}}T1,\\ FLAIR\end{tabular}} &
  \multicolumn{1}{l|}{\begin{tabular}[c]{@{}l@{}}Brain atrophy per\\ region, WMH, and\\ cortical thickness.\\ Brain structures\\ segmented include \\ 2 vessels.\end{tabular}} &
  \multicolumn{1}{l|}{DL} &
  \multicolumn{1}{l|}{FDA (II), CE (IIa)} \\ \hline
  
\multicolumn{1}{|l|}{Heuron} &
  \multicolumn{1}{l|}{\begin{tabular}[c]{@{}l@{}}StroCare\\ Suite,\\ AgingCare\\ Suite\end{tabular}} &
  \multicolumn{1}{l|}{\begin{tabular}[c]{@{}l@{}}ICH, LVO,\\ ISL,\\ Atrophy\end{tabular}} &
  \multicolumn{1}{l|}{\begin{tabular}[c]{@{}l@{}}NCCT,\\ CTA,\\ CTP,\\ T1\end{tabular}} &
  \multicolumn{1}{l|}{\begin{tabular}[c]{@{}l@{}}ICH detection and \\ triage from NCCT.\\ LVO detection and \\ hemisphere from\\ NCCT. ASPECTS\\ and regional analy-\\ sis from NCCT. \\ LVO detection and \\ hemisphere from \\ CTA. Volume mis-\\ match from CTP. \\ Atrophy per region \\ from T1.\end{tabular}} &
  \multicolumn{1}{l|}{AI} &
  \multicolumn{1}{l|}{\begin{tabular}[c]{@{}l@{}}FDA (II)\\ (StroCare Suite \\ ICH)\end{tabular}} \\ \hline

\multicolumn{1}{|l|}{Quibim} &
  \multicolumn{1}{l|}{QP-Brain} &
  \multicolumn{1}{l|}{\begin{tabular}[c]{@{}l@{}}WMH,\\ Atrophy\end{tabular}} &
  \multicolumn{1}{l|}{\begin{tabular}[c]{@{}l@{}}T1,\\ FLAIR\end{tabular}} &
  \multicolumn{1}{l|}{\begin{tabular}[c]{@{}l@{}}Brain atrophy per \\ region from T1. \\ Segmentation and \\ volume of WMH \\ and longitudinal \\ analysis from \\ FLAIR.\end{tabular}} &
  \multicolumn{1}{l|}{AI, DL} &
  \multicolumn{1}{l|}{FDA (II), CE (IIb)} \\ \hline

\multicolumn{1}{|l|}{QYNAPSE} &
  \multicolumn{1}{l|}{QyScore} &
  \multicolumn{1}{l|}{\begin{tabular}[c]{@{}l@{}}WMH,\\ Atrophy\end{tabular}} &
  \multicolumn{1}{l|}{\begin{tabular}[c]{@{}l@{}}T1,\\ FLAIR\end{tabular}} &
  \multicolumn{1}{l|}{\begin{tabular}[c]{@{}l@{}}Brain atrophy per \\ region from T1. \\ Segmentation and \\ volume of WMH \\ from T1 and FLAIR.\end{tabular}} &
  \multicolumn{1}{l|}{AI} &
  \multicolumn{1}{l|}{FDA (II), CE (IIa)} \\ \hline

\multicolumn{1}{|l|}{\begin{tabular}[c]{@{}l@{}}United Imaging\\ Intelligence\end{tabular}} &
  \multicolumn{1}{l|}{\begin{tabular}[c]{@{}l@{}}uAI\\ Discover\end{tabular}} &
  \multicolumn{1}{l|}{\begin{tabular}[c]{@{}l@{}}ICH, ISL,\\ WMH,\\ Atrophy,\\ PVS,\\ Lacunes,\\ CMB, BA\end{tabular}} &
  \multicolumn{1}{l|}{\begin{tabular}[c]{@{}l@{}}NCCT,\\ CTA,\\ CTP,\\ MRI\end{tabular}} &
  \multicolumn{1}{l|}{\begin{tabular}[c]{@{}l@{}}Segmentation of \\ ICH, edema, and \\ midline shift from\\ \\ NCCT. ASPECTS \\ score per region\\ from NCCT. \\ Detection of BA\\ from CTA. Segmen-\\ tation of infarct core\\ and mismatch areas\\ from CTP. Quantit-\\ ative evaluation of \\ ISL, lacunes, WMH,\\ PVS, CMB, and\\ atrophy from MRI to\\ provide a total\\ CSVD score.\end{tabular}} &
  \multicolumn{1}{l|}{AI} &
  \multicolumn{1}{l|}{-} \\ \hline
  
\end{longtable}
\normalsize

\clearpage
\subsection{Research systems}

\textbf{Table \ref{tab:papers1}} shows an overview of the 13 research publications identified, with more detail provided in \textbf{Table \ref{tab:papers2}}. Only 1 publication focuses on object detection, with the remainder focusing on segmentation. Below we summarise the types of CVD markers and imaging modalities targeted in the publications, the data characteristics, the image analysis methods employed, the validation methods and results, and the risk of bias.

\subsubsection{CVD markers and imaging modalities}

\begin{figure}
    \begin{subfigure}{0.49\textwidth}
        \centering
        \includegraphics[width=\textwidth]{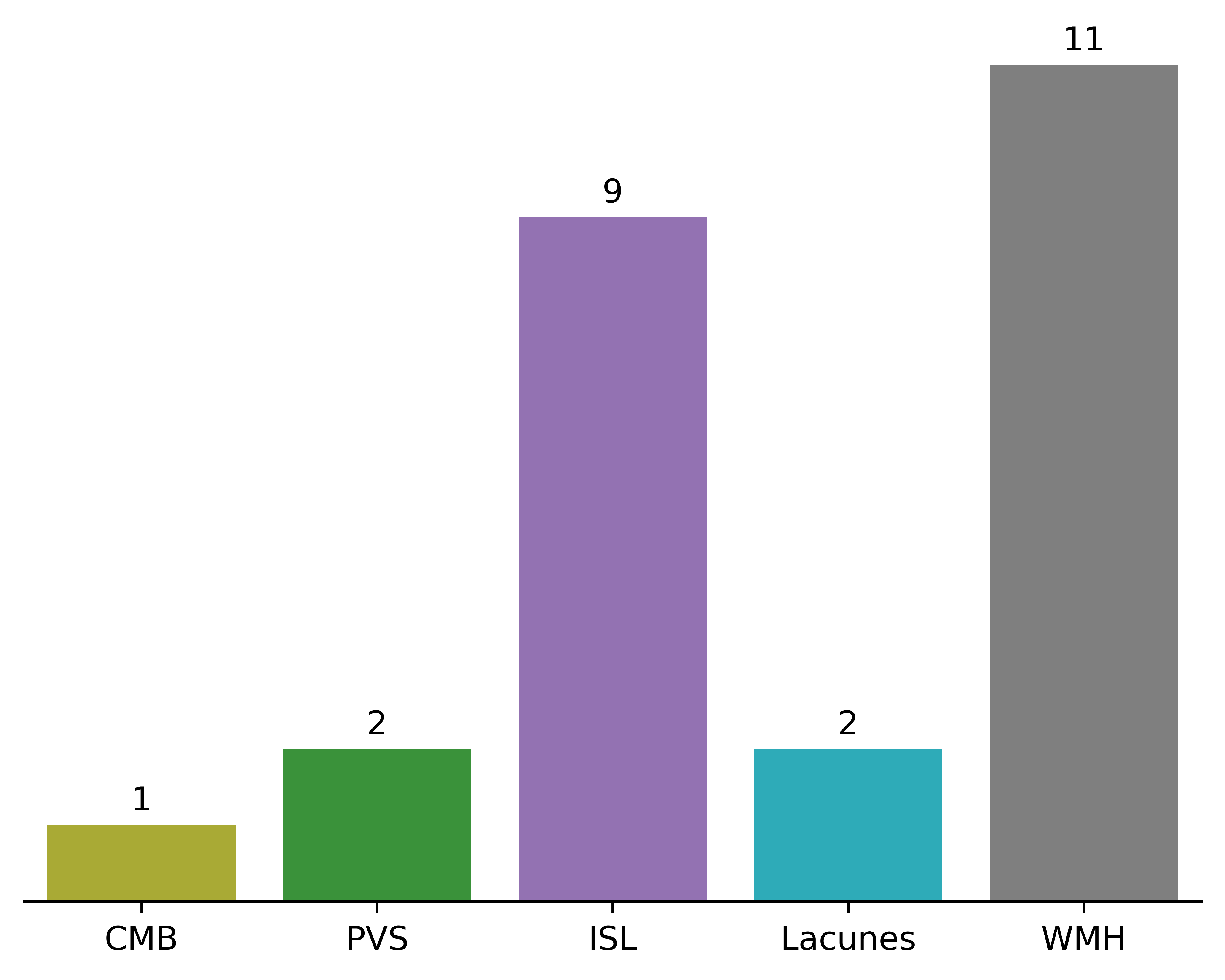}
        \caption{Number of publications targeting each pathology}
        \label{fig:path dist research bar}
    \end{subfigure}
    \hfill
    \begin{subfigure}{0.49\textwidth}
        \centering
        \includegraphics[width=\textwidth]{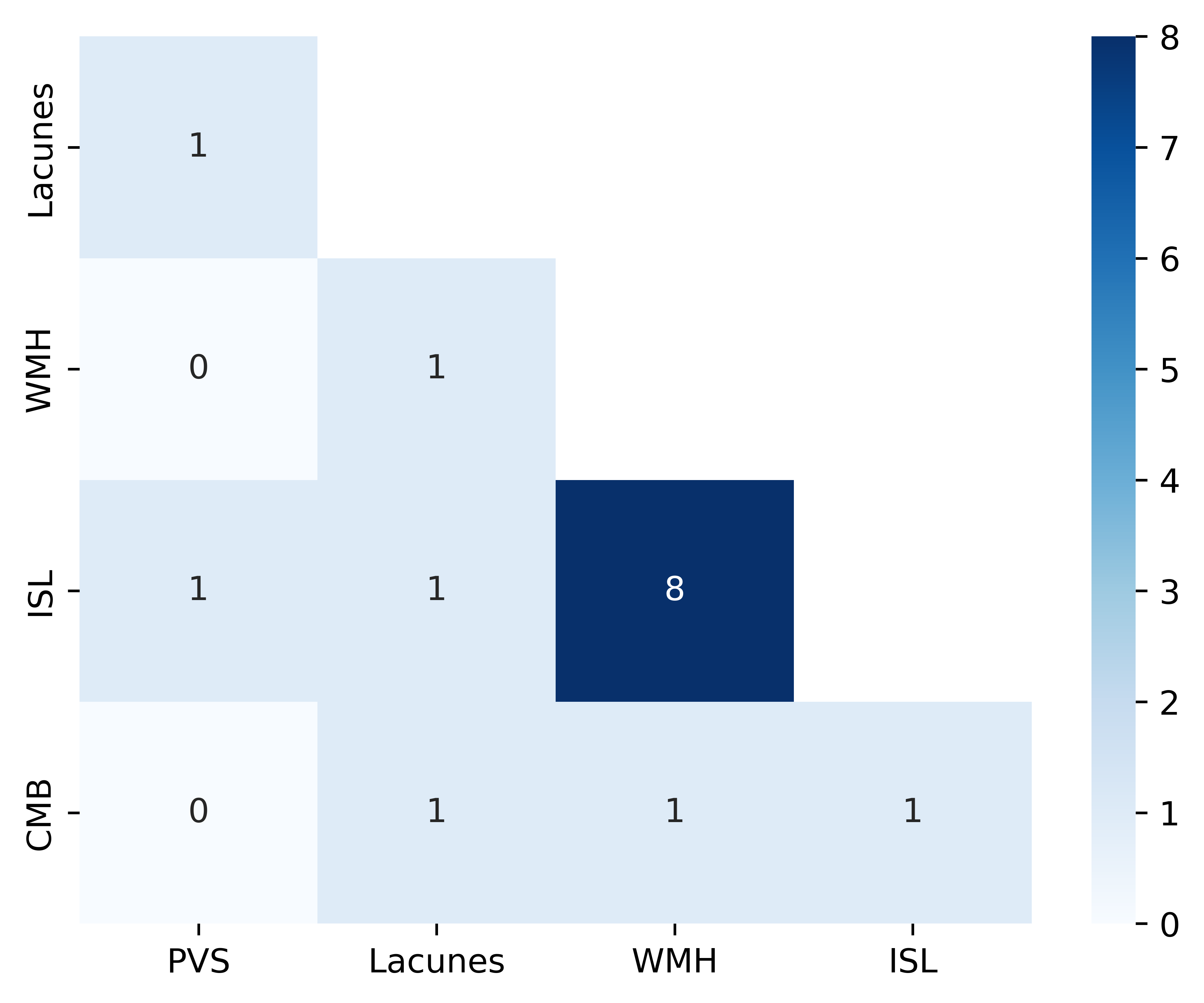}
        \caption{Co-occurrence of pathologies targeted}
        \label{fig:path dist research matrix}
    \end{subfigure}
    \hfill
    \begin{subfigure}{0.49\textwidth}
        \centering
        \includegraphics[width=\textwidth]{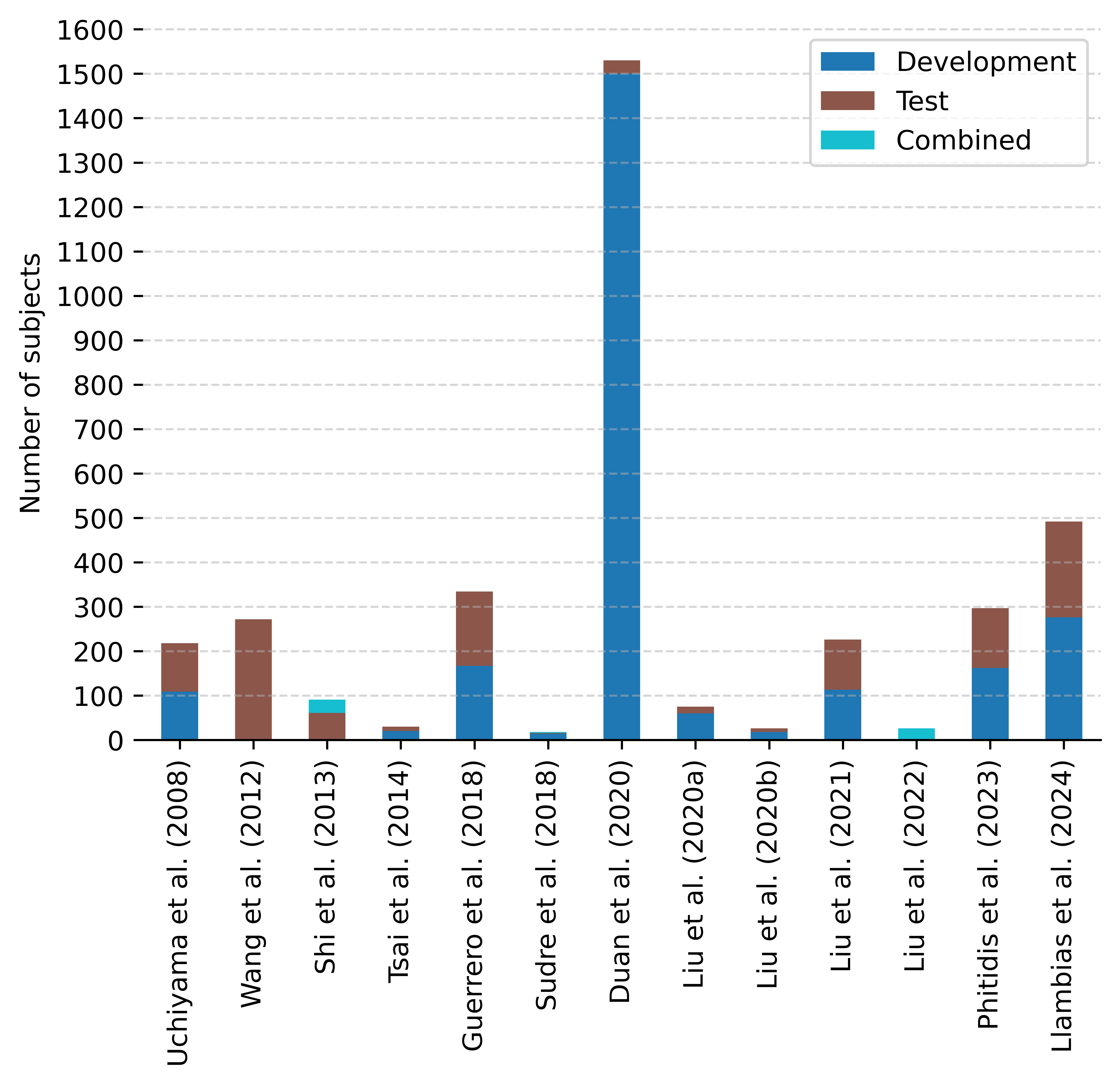}
        \caption{Dataset size}
        \label{fig:dataset size research}
    \end{subfigure}
    \hfill
    \begin{subfigure}{0.49\textwidth}
        \centering
        \includegraphics[width=\textwidth]{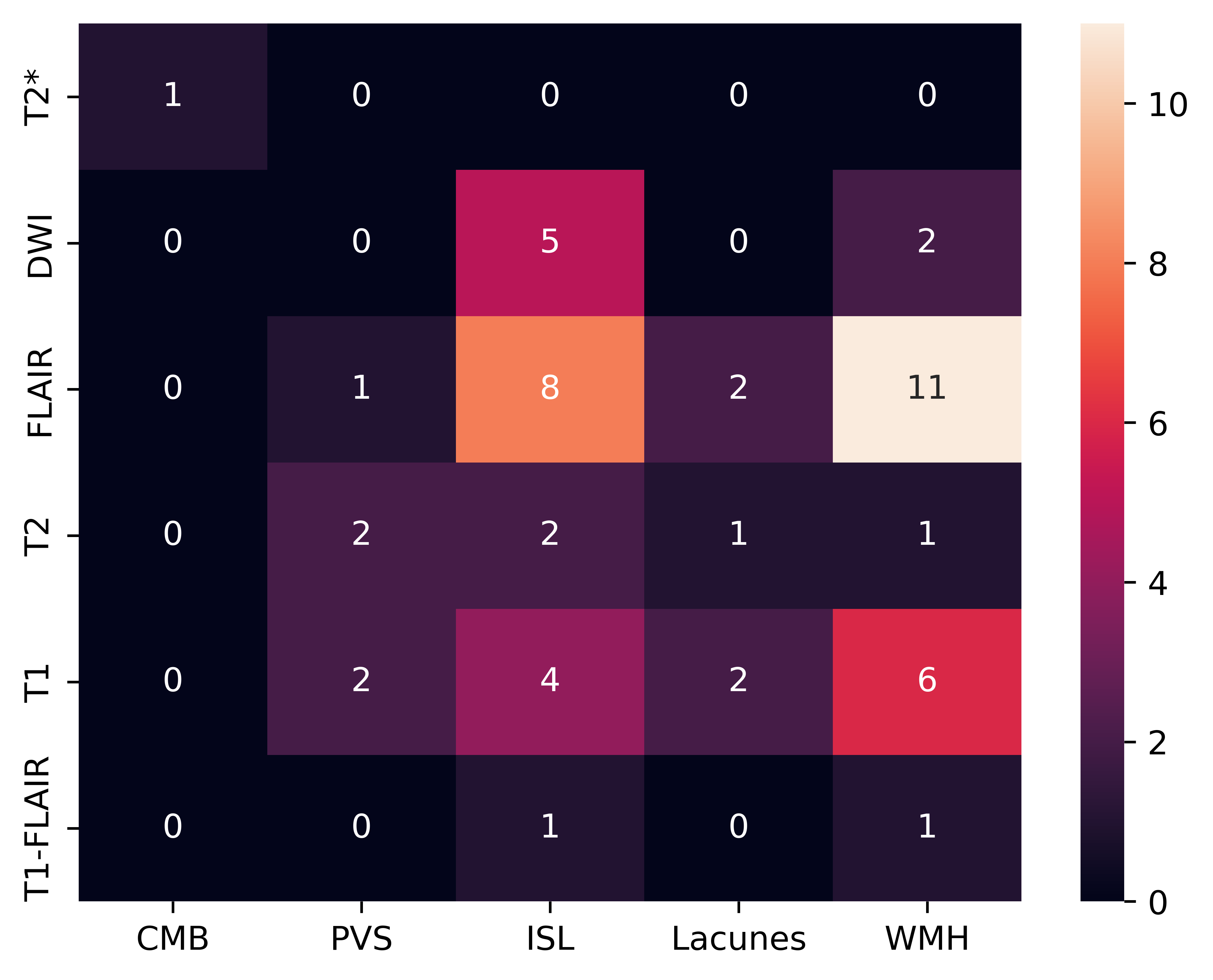}
        \caption{Number of uses of each modality for each pathology}
        \label{fig:path mod research}
    \end{subfigure}
    \caption{Pathology, modality and sample size characteristics of identified research publications.}
    \label{fig:characteristics research}
\end{figure}

Some CVD markers were more heavily targeted than others: we see in \textbf{Fig \ref{fig:path dist research bar}} that 11/13 publications dealt with WMH, 9/13 with ISL, 2/13 with PVS, 2/13 with lacunes, and 1/13 with CMB. 3 papers aimed to segment WMH in the presence of ISL (without attempting to segment ISL) \citep{shi_automated_2013,tsai_automated_2014,liu_deep_2020}. \textbf{Fig \ref{fig:path dist research matrix}} shows the co-occurrence of targeted CVD markers within the identified publications. This reveals that the only CVD markers which are targeted jointly more than once are WMH and ISL.

In terms of imaging modalities, \textbf{Fig \ref{fig:path mod research}} shows the number of uses of each MRI sequence type to identify (or identify confounders of) each CVD feature. Of the 13 papers, 11 used at least 2 sequences. The only paper to tackle CMB segmentation \citep{duan_primary_2020} used T2*, which is essential for identifying these very small bleeds. For tackling PVS, both papers \citep{uchiyama_computer-aided_2008, sudre_3d_2018} used T1 and T2 while the latter additionally used FLAIR. For tackling ISL, the most popular sequences were T1, FLAIR and DWI, with 2 papers also utilising T2. 3 papers exclusively use DWI for ISL identification \citep{shi_automated_2013,tsai_automated_2014,duan_primary_2020}. For tackling lacunes, both \citet{sudre_3d_2018} and \citet{duan_primary_2020} used T1 and FLAIR, with the former also using T2. For tackling WMH, all papers use FLAIR with some additionally using T1. Overall, FLAIR was used the most times to identify a CVD marker, with 22 usages, T1 was used 14 times, T2 was used 6 times, DWI was used 5 times and T2* and T1-FLAIR were each used only once.

\subsubsection{Sample characteristics}

\paragraph{Data acquisition} Non-open source data was used by 10/13 papers. The remaining 3 studies \citep{liu_deep_2020,liu_attention_2020, llambias2024heterogeneous}, used publicly available data from the WMH segmentation challenge\footnote{\url{https://wmh.isi.uu.nl/}} \citep{kuijf_standardized_2019}, the ISLES challenge\footnote{\url{https://isles22.grand-challenge.org/}} \citep{hernandez_petzsche_isles_2022}, the MSSEG challenge \citep{commowick2021multiple}, and the MS08 challenge \citep{styner20083d}. \citet{liu_deep_2020} and \citet{liu_attention_2020} added additional lesion annotations of their own.

\paragraph{Reference standard} In 7/13 studies, multiple expert annotators generated or endorsed the ground truth \citep{wang_multi-stage_2012,sudre_3d_2018,duan_primary_2020,liu_deep_2020,liu_attention_2020, liu2021deep, liu2022llrhnet}, with the remaining 6/13 studies using annotations generated by a single expert. Most studies used manually generated ground truth, however 3/13 studies state that the ground truth was semi-automatically generated \citep{tsai_automated_2014,guerrero_white_2018, phitidis2023segmentation}. All identified publications except \citet{llambias2024heterogeneous} utilised data where, for at least some subjects, multiple CVD features are visible within the same image.

\paragraph{Demographics} Only 5/13 papers \citep{wang_multi-stage_2012,shi_automated_2013,tsai_automated_2014, liu2021deep, phitidis2023segmentation} provided explicit information regarding the mean age and sex of the subjects in their dataset. The mean age ranged from 52 to 70 and comprised a total of 466 male and 337 female subjects.

\paragraph{Data splits} Data splits into development/training and test sets can be seen in \textbf{Fig \ref{fig:dataset size research}}. In cases where K-fold cross validation was used, the development and test sets are considered to be the total dataset size.
For methods which do not use machine learning, 2/3 clearly utilise an independent test set \citep{shi_automated_2013,tsai_automated_2014}, whilst \citet{wang_multi-stage_2012} do not specify their development data, only their test data.
6/10 machine learning approaches \citep{uchiyama_computer-aided_2008,guerrero_white_2018,sudre_3d_2018,duan_primary_2020,liu_deep_2020,liu_attention_2020, phitidis2023segmentation, llambias2024heterogeneous} utilise a separate test set and 3/10 perform K-fold cross validation \citep{uchiyama_computer-aided_2008,guerrero_white_2018, liu2021deep}. \citet{liu2022llrhnet} do not make clear the train/test split used. With the exception of \citep{duan_primary_2020}, no publications use more than 297 subjects in total, for development and testing.

\subsubsection{Preprocessing}

To handle the variation in appearance between scans acquired from different scanners, different imaging protocols, and different subjects, some preprocessing is usually performed to normalise the data before running the core image analysis detection/segmentation algorithm. 
In machine learning, preprocessing is easily defined: it comprises the steps taken before inputs are passed into the machine learning model. These steps should be performed in the same way for the training and testing data. For analytical methods which rely entirely on rules-based processing, it can be more difficult to identify where preprocessing ends and the algorithm begins. In this review, we consider preprocessing to be conventional steps which could feasibly be applied prior to any image processing algorithm, such as registration to a standard template, or normalisation. We consider, for example, more complex tasks like unsupervised brain tissue segmentation as part of the method, not the preprocessing.

\citet{liu_deep_2020}, \citet{liu_attention_2020}, and \citet{llambias2024heterogeneous} use publicly available datasets, without applying any further preprocessing.
We now describe the preprocessing steps taken in the remaining 9 publications (\citet{uchiyama_computer-aided_2008} do not mention preprocessing), making use of non-open source data. 

\paragraph{Bias field correction} Bias field correction was applied by 4/9 papers \citep{wang_multi-stage_2012,sudre_3d_2018, liu2022llrhnet, phitidis2023segmentation}. \citet{wang_multi-stage_2012} specified the method as non-parametric non-uniform intensity normalisation \citep{sled_nonparametric_1998} and \citet{liu2022llrhnet} used a variant proposed in \citet{tustison2010n4itk}.

\paragraph{Coregistration} Coregistration of multiple sequences is stated by 5/9 \citep{wang_multi-stage_2012,shi_automated_2013,tsai_automated_2014,guerrero_white_2018, liu2022llrhnet}. Of these 5 papers, 2 \citep{wang_multi-stage_2012,guerrero_white_2018} specify the use of FSL-FLIRT\footnote{\url{https://fsl.fmrib.ox.ac.uk/fsl/fslwiki/FLIRT}} \citep{jenkinson_global_2001,jenkinson_improved_2002} as the software used to perform affine registration. Additionally, 1 paper specifies the transformation as affine \citep{shi_automated_2013} and 1 specifies rigid \citep{tsai_automated_2014}. It should be noted that rigid transformation is a subset of affine transformation for which scaling and shearing is not allowed. 
The cost function is specified as normalised mutual information \citep{maes_multimodality_1997} by 2 papers \citep{shi_automated_2013,tsai_automated_2014} and as cross-correlation by \citet{wang_multi-stage_2012}. None specify the optimisation algorithm or the interpolation method. 

\paragraph{Brain extraction} Brain extraction may be equivalently referred to as skull stripping. Brain extraction was reported by 5/9 papers \citep{wang_multi-stage_2012,tsai_automated_2014,sudre_3d_2018, liu2022llrhnet, phitidis2023segmentation}, with the former 2 using FSL-BET\footnote{\url{https://fsl.fmrib.ox.ac.uk/fsl/fslwiki/BET}} \citep{smith_fast_2002} and \citet{liu2022llrhnet} using AFNI \citep{cox1996afni}. It may be additionally suspected that \citet{guerrero_white_2018} performed skull stripping, based on images shown in the paper.

\paragraph{Resampling} Only \citet{guerrero_white_2018} state that they resample images to a specific spatial resolution (1 mm in the axial plane). \citet{phitidis2023segmentation} utilise nnU-Net \citep{isensee_automated_2019} for selecting the spatial resolution.

\paragraph{Normalisation} 5/9 papers state that they perform normalisation or standardisation (z-scoring) of voxel intensities \citep{shi_automated_2013,tsai_automated_2014,guerrero_white_2018,sudre_3d_2018,duan_primary_2020}. Of these, 3 focused on the white matter region; 1 study \citep{tsai_automated_2014} scaled the voxels in this region to the range [0,100], 1 study \citep{sudre_3d_2018} standardised the image using the white matter statistics, and 1 study \citep{duan_primary_2020} normalised and aligned the histogram peaks based on the white matter region. \citet{guerrero_white_2018} standardised to the non-background statistics and clipped values below 3 standard deviations from the mean. \citet{shi_automated_2013} simply states that they performed intensity normalisation. Additionally, \citet{liu2021deep} perform a gamma transformation on the image patches extracted from the first stage of their solution before passing them to the second stage, in order to stretch the high intensity regions. \citet{phitidis2023segmentation} used the intensity normalisation scheme of nnU-Net.

\subsubsection{Rule-based and machine learning methods}

The following studies used rules-based (RB) and classical machine learning (ML) methods:

\paragraph{\citet{uchiyama_computer-aided_2008}} The authors used the morphological white top-hat transform to enhance hyperintense lesions in the T2 sequences and then applied thresholding to segment the regions of interest. They extracted features including size, shape, location, and difference in T1 and T2 signal intensity. They then used these features to train a 3 layer neural network to classify the segmented lesion as either PVS or lacunar infarct (the result of an ischaemic stroke affecting blood supply to the deep brain tissue). In later work \citep{uchiyama_cad_2009}, they introduced the idea of registering a magnetic resonance angiography (MRA) image to the T1 and T2 images in order to use blood flow to differentiate between lacunar infarcts and PVS, although this was not validated.

\paragraph{\citet{wang_multi-stage_2012}} Following preprocessing, the authors used Freesurfer\footnote{\url{https://surfer.nmr.mgh.harvard.edu/}} to segment brain tissues from the T1 images and followed this with a combination of image processing and computer vision techniques on the T1, T2 and FLAIR scans. For instance, they fit Gaussian mixture models (GMMs) to identify relevant tissue thresholds prior to thresholding, then apply morphological operations of erosion, dilation and region growing. 

\paragraph{\citet{shi_automated_2013}} The authors segmented WMH in the presence of acute infarcts. Following preprocessing, they fused the T1 and FLAIR images and from this segmented white matter, grey matter and cerebrospinal fluid. To identify the WMH they focused on FLAIR and performed: grayscale morphological closing followed by a local standard deviation operation to identify boarders of hyperintense regions; thresholding; binary morphological dilation; grayscale morphological reconstruction within the mask area; thresholding the difference image between the original FLAIR and reconstructed image; and finally binary morphological reconstruction. Infarcts were detected in DWI using thresholding, dilation, and morphological reconstruction and then subtracted from the final WMH segmentation.  

\paragraph{\citet{tsai_automated_2014}} Following preprocessing, the authors developed a scheme based on fusion of T1 and FLAIR images. Probabilistic white and grey matter masks were generated using the New Segment module of SPM8\footnote{\url{https://www.fil.ion.ucl.ac.uk/spm/}} on the fused image and a brain template (ICBM), and these were used to estimate the registration between the images with the DARTEL module of SPM8. In the same space, a logical AND operation was performed between appropriate template labels and the binarised probabilistic white matter mask of the image, resulting in the final white matter mask. The FLAIR image was normalised within this region and thresholded to located candidate WMH. A mask of the grey/white matter junction was generated from the fused image and WMH which significantly overlapped with this mask were eliminated. Finally ISL were detected by thresholding the DWI image and WMH overalpping these were eliminated also.

\subsubsection{Deep learning methods}
A summary of the design choices made by the methods that used DL can be seen in \textbf{Table \ref{tab:DL research}}. Some common themes are the use of CNNs, cross-entropy or Dice loss, batch normalisation, and the ReLu activation function. The following studies used DL:

\paragraph{\citet{guerrero_white_2018}} The authors implemented a 2D CNN based on the successful U-Net architecture \citep{ronneberger_u-net_2015}, which they termed uResNet. Their main enhancement of the standard U-Net model was introducing residual connections \citep{he_deep_2016} within each layer.

\paragraph{\citet{sudre_3d_2018}} A 3D model based on RCNN \citep{girshick_rich_2014} was developed by the authors to detect lacunes and PVS. They do not resize or pool the selected bounding box regions as is done in the original framework, since the target objects (lacunes and PVS) are small making them prone to distortion during resizing, likely leading to worse model performance. Their model attempts to classify each region as nothing, lacune, PVS, or either lacune or PVS.

\paragraph{\citet{duan_primary_2020}} The authors used 4 independently trained 2D U-Nets to segment WMH, lacunes, infarcts, and microbleeds from different sequences as detailed in \textbf{Table \ref{tab:papers2}}.

\paragraph{\citet{liu_deep_2020}} The authors developed a 2D CNN method which they coined M2DCNN. The model utilised 2 symmetric U-Net-like networks at different resolutions to extract multi-scale features. At the encoder-decoder bottlenecks they employed dilated convolutions with a stride of 1 to enlarge the receptive field size whilst maintaining the same feature map dimensions. They also replaced the standard convolution blocks with dense blocks \citep{huang_densely_2017}.

\paragraph{\citet{liu_attention_2020}} A 2D U-Net model was developed by the authors, termed DRANet, incorporating an attention block acting on the residual features. This block has 2 internal branches, comprising the trunk branch which is made up of residual blocks and the dilated soft mask branch which incorporates dilated convolutions and ends with a sigmoid activation. The output of the two branches are then summed element-wise. The authors claim that since the inputs to the trunk and soft mask branches are the same, then the soft mask branch will either construct the identity mapping to reinforce the residual features, or contribute new features to suppress noise which may be present.

\paragraph{\citet{liu2021deep}} The authors implemented a two stage approach for segmenting small focal cerebral ischaemia and lacunar infarcts. The first stage utilised T2-FLAIR input to a 2D U-Net producing a binary mask to differentiate foreground lesions from background. For each predicted foreground connected component, a 32 $\times$ 32 crop of the corresponding T1-FLAIR images was taken about its centroid and used as input to a CNN which classified the lesion.

\paragraph{\citet{liu2022llrhnet}} The authors modify a 2D U-Net to include a global branch utilising transformer layers. The self attention mechanism working on flattened image patches allows long-range features to be learned with even a single layer, which is not the case for convolution. The features are incorporated at the bottleneck of the U-Net. 

\paragraph{\citet{phitidis2023segmentation}} In this work, the authors evaluated two existing self-configuring, out-of-the-box segmentation frameworks: nnU-Net \citep{isensee2021nnu}, and Auto3DSeg\footnote{\url{https://monai.io/apps/auto3dseg}} \citep{monai2020monai}, for the WMH and ISL segmentation task, from FLAIR images.

\paragraph{\citet{llambias2024heterogeneous}} The authors trained a model to segment WMH, ISL, and MS lesions using separate, partially labelled datasets. They initially trained a 2D U-Net model on all the datasets to perform binary lesion segmentation. Then, they fine-tuned the model to perform multiclass segmentation, using a weighted sampling strategy to ensure an equal number of slices with each pathology. Then, the further fine-tuned the model, using the same sampling strategy, but reverting back to the binary segmentation setting. To classify the lesion type, they trained a fully connected network (FCN) on manually extracted features of each connected component. These features were the volume, surface area, orientation, and principle axis length.

\begin{table}[]
\centering
\caption{Deep learning system design choices for included publications.}
\label{tab:DL research}
\resizebox{\textwidth}{!}{%
\begin{tabular}{llllllllllllll}
\hline
\multicolumn{1}{|l|}{\textbf{Study}} &
  \multicolumn{1}{l|}{\textbf{Pre\textsuperscript{1}}} &
  \multicolumn{1}{l|}{\textbf{Arch\textsuperscript{2}}} &
  \multicolumn{1}{l|}{\textbf{Post\textsuperscript{3}}} &
  \multicolumn{1}{l|}{\textbf{Aug\textsuperscript{4}}} &
  \multicolumn{1}{l|}{\textbf{Loss\textsuperscript{5}}} &
  \multicolumn{1}{l|}{\textbf{Dim}} &
  \multicolumn{1}{l|}{\textbf{Dil}} &
  \multicolumn{1}{l|}{\textbf{Norm\textsuperscript{6}}} &
  \multicolumn{1}{l|}{\textbf{Act}} &
  \multicolumn{1}{l|}{\textbf{Drop}} &
  \multicolumn{1}{l|}{\textbf{MS}} &
  \multicolumn{1}{l|}{\textbf{LR}} &
  \multicolumn{1}{l|}{\textbf{Ens}} \\ \hline
\multicolumn{1}{|l|}{\citet{guerrero_white_2018}} &
  \multicolumn{1}{l|}{K,R,S} &
  \multicolumn{1}{l|}{uResNet} &
  \multicolumn{1}{l|}{} &
  \multicolumn{1}{l|}{F} &
  \multicolumn{1}{l|}{CE} &
  \multicolumn{1}{l|}{2D} &
  \multicolumn{1}{l|}{} &
  \multicolumn{1}{l|}{B} &
  \multicolumn{1}{l|}{ReLu} &
  \multicolumn{1}{l|}{} &
  \multicolumn{1}{l|}{} &
  \multicolumn{1}{l|}{\checkmark} &
  \multicolumn{1}{l|}{} \\ \hline
\multicolumn{1}{|l|}{\citet{sudre_3d_2018}} &
  \multicolumn{1}{l|}{B,K,S} &
  \multicolumn{1}{l|}{RCNN} &
  \multicolumn{1}{l|}{} &
  \multicolumn{1}{l|}{} &
  \multicolumn{1}{l|}{\begin{tabular}[c]{@{}l@{}}CE,\\ RMSE\end{tabular}} &
  \multicolumn{1}{l|}{3D} &
  \multicolumn{1}{l|}{\checkmark} &
  \multicolumn{1}{l|}{B} &
  \multicolumn{1}{l|}{ReLu} &
  \multicolumn{1}{l|}{} &
  \multicolumn{1}{l|}{} &
  \multicolumn{1}{l|}{} &
  \multicolumn{1}{l|}{} \\ \hline
\multicolumn{1}{|l|}{\citet{duan_primary_2020}} &
  \multicolumn{1}{l|}{H} &
  \multicolumn{1}{l|}{U-Net} &
  \multicolumn{1}{l|}{R} &
  \multicolumn{1}{l|}{} &
  \multicolumn{1}{l|}{} &
  \multicolumn{1}{l|}{2D} &
  \multicolumn{1}{l|}{} &
  \multicolumn{1}{l|}{} &
  \multicolumn{1}{l|}{} &
  \multicolumn{1}{l|}{} &
  \multicolumn{1}{l|}{} &
  \multicolumn{1}{l|}{} &
  \multicolumn{1}{l|}{\checkmark} \\ \hline
\multicolumn{1}{|l|}{\citet{liu_deep_2020}} &
  \multicolumn{1}{l|}{} &
  \multicolumn{1}{l|}{M2DCNN} &
  \multicolumn{1}{l|}{} &
  \multicolumn{1}{l|}{} &
  \multicolumn{1}{l|}{D*} &
  \multicolumn{1}{l|}{2D} &
  \multicolumn{1}{l|}{\checkmark} &
  \multicolumn{1}{l|}{B} &
  \multicolumn{1}{l|}{ReLu} &
  \multicolumn{1}{l|}{\checkmark} &
  \multicolumn{1}{l|}{\checkmark} &
  \multicolumn{1}{l|}{} &
  \multicolumn{1}{l|}{} \\ \hline
\multicolumn{1}{|l|}{\citet{liu_attention_2020}} &
  \multicolumn{1}{l|}{} &
  \multicolumn{1}{l|}{DRANet} &
  \multicolumn{1}{l|}{} &
  \multicolumn{1}{l|}{} &
  \multicolumn{1}{l|}{D} &
  \multicolumn{1}{l|}{2D} &
  \multicolumn{1}{l|}{\checkmark} &
  \multicolumn{1}{l|}{} &
  \multicolumn{1}{l|}{} &
  \multicolumn{1}{l|}{\checkmark} &
  \multicolumn{1}{l|}{\checkmark} &
  \multicolumn{1}{l|}{} &
  \multicolumn{1}{l|}{} \\ \hline
\multicolumn{1}{|l|}{\citet{liu2021deep}} &
  \multicolumn{1}{l|}{} &
  \multicolumn{1}{l|}{\begin{tabular}[c]{@{}l@{}}U-Net, \\ CNN\end{tabular}} &
  \multicolumn{1}{l|}{} &
  \multicolumn{1}{l|}{R} &
  \multicolumn{1}{l|}{CE} &
  \multicolumn{1}{l|}{2D} &
  \multicolumn{1}{l|}{} &
  \multicolumn{1}{l|}{B} &
  \multicolumn{1}{l|}{ReLu} &
  \multicolumn{1}{l|}{Y} &
  \multicolumn{1}{l|}{} &
  \multicolumn{1}{l|}{} &
  \multicolumn{1}{l|}{\checkmark} \\ \hline
\multicolumn{1}{|l|}{\citet{liu2022llrhnet}} &
  \multicolumn{1}{l|}{B,K,R} &
  \multicolumn{1}{l|}{LLRHNet} &
  \multicolumn{1}{l|}{} &
  \multicolumn{1}{l|}{} &
  \multicolumn{1}{l|}{D} &
  \multicolumn{1}{l|}{2D} &
  \multicolumn{1}{l|}{} &
  \multicolumn{1}{l|}{L} &
  \multicolumn{1}{l|}{ReLu} &
  \multicolumn{1}{l|}{\checkmark} &
  \multicolumn{1}{l|}{\checkmark} &
  \multicolumn{1}{l|}{} &
  \multicolumn{1}{l|}{} \\ \hline
\multicolumn{1}{|l|}{\citet{llambias2024heterogeneous}} &
  \multicolumn{1}{l|}{} &
  \multicolumn{1}{l|}{\begin{tabular}[c]{@{}l@{}}U-Net,\\ FCN\end{tabular}} &
  \multicolumn{1}{l|}{} &
  \multicolumn{1}{l|}{\begin{tabular}[c]{@{}l@{}}F,R,\\ E,N,\\ B,M\end{tabular}} &
  \multicolumn{1}{l|}{D, CE} &
  \multicolumn{1}{l|}{2D} &
  \multicolumn{1}{l|}{} &
  \multicolumn{1}{l|}{} &
  \multicolumn{1}{l|}{} &
  \multicolumn{1}{l|}{} &
  \multicolumn{1}{l|}{} &
  \multicolumn{1}{l|}{\checkmark} &
  \multicolumn{1}{l|}{\checkmark} \\ \hline
\multicolumn{14}{l}{\begin{tabular}[c]{@{}l@{}}1. Preprocessing: B = bias field correction, H = histogram based normalisation, K = skull stripping, N = normalisation \\ (scaling), R = resampling, S = standardisation (z-score), T = transformation to template (e.g.~NMI).\\ 2. Architecture.\\ 3. Postprocessing: R = mask subtraction rules.\\ 4. Augmentation: F = flipping, R = rotation, E = elastic, N = noise, B = bias field, M = motion artefacts.\\ 5. Loss function: CE = cross-entropy, D = Dice, RMSE: root mean square error.\\ 6. Normalisation: B = batch, I = instance, L = Layer.\\ Act: activation function, Dim: dimension of convolutions, Dil: use of dilated convolutions, Drop: use of dropout, \\ MS: use of multi-scale features (processed separately), LR: use of learning rate scheduler, Ens: ensemble of models.\\ Note: \citet{phitidis2023segmentation} not included since the authors use existing frameworks.\\ $*$ Variant of standard dice loss where the function is: 1 - lesion\_dice\_score - background\_dice\_score\end{tabular}}
\end{tabular}%
}
\end{table}

\subsubsection{Validation}

From \textbf{Fig \ref{fig:dataset size research}} it can be seen that of the papers which gave information regarding their development and testing data splits \citep{uchiyama_computer-aided_2008,shi_automated_2013,tsai_automated_2014,guerrero_white_2018,sudre_3d_2018,duan_primary_2020,liu_deep_2020,liu_attention_2020, liu2021deep, phitidis2023segmentation, llambias2024heterogeneous}, 5/10 studies \citep{uchiyama_computer-aided_2008,shi_automated_2013,tsai_automated_2014,guerrero_white_2018} utilised at least 50\% of their total data for testing. Notably, \citet{duan_primary_2020} used only 30 cases (with 2-4 CVD markers) for testing, despite the comparatively large amount of training data.

\begin{figure}[!htb]
    \begin{subfigure}{0.49\textwidth}
        \centering
        \includegraphics[width=\textwidth]{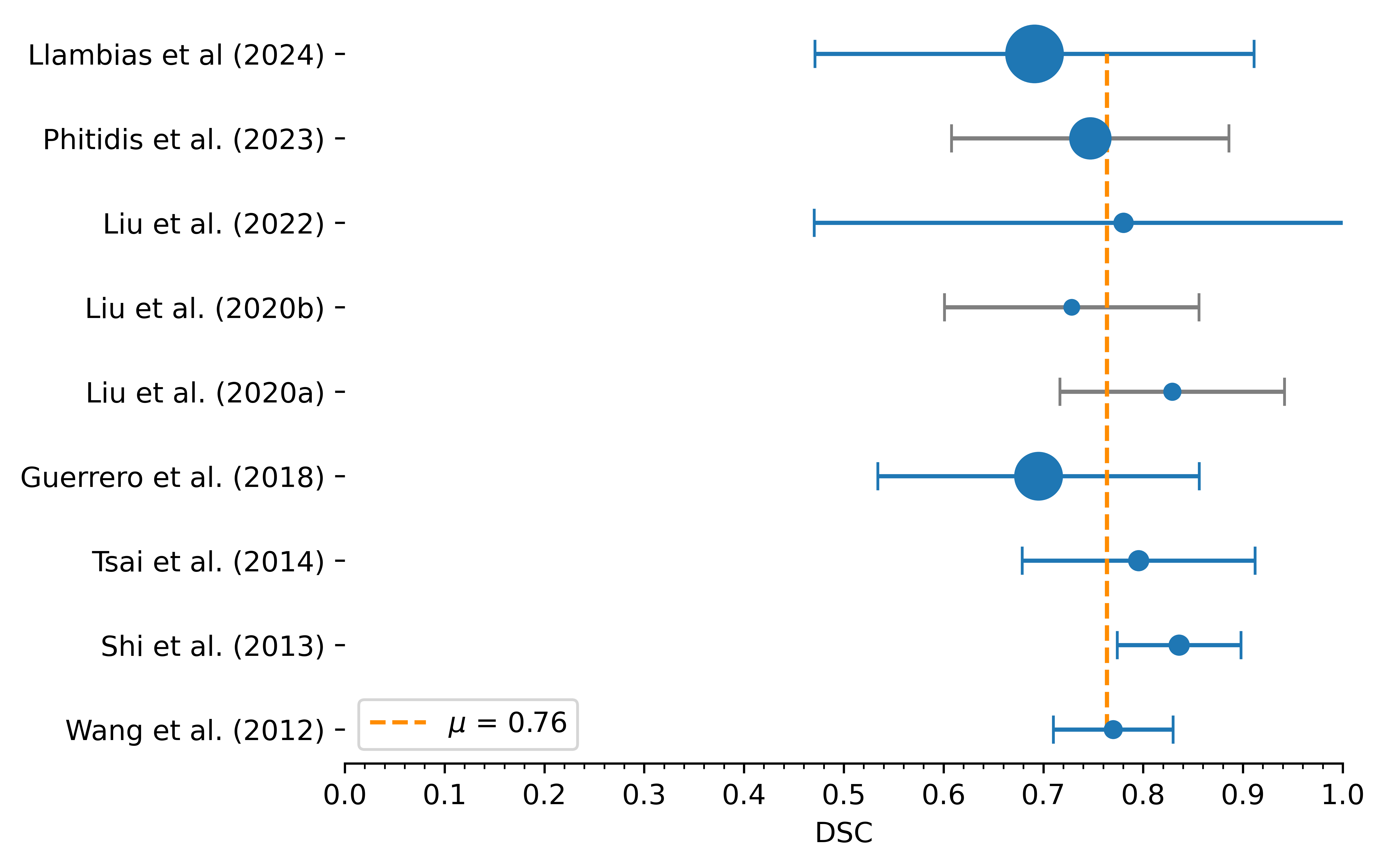}
        \caption{WMH}
        \label{fig:foreset WMH research}
    \end{subfigure}
    \hfill
    \begin{subfigure}{0.49\textwidth}
        \centering
        \includegraphics[width=\textwidth]{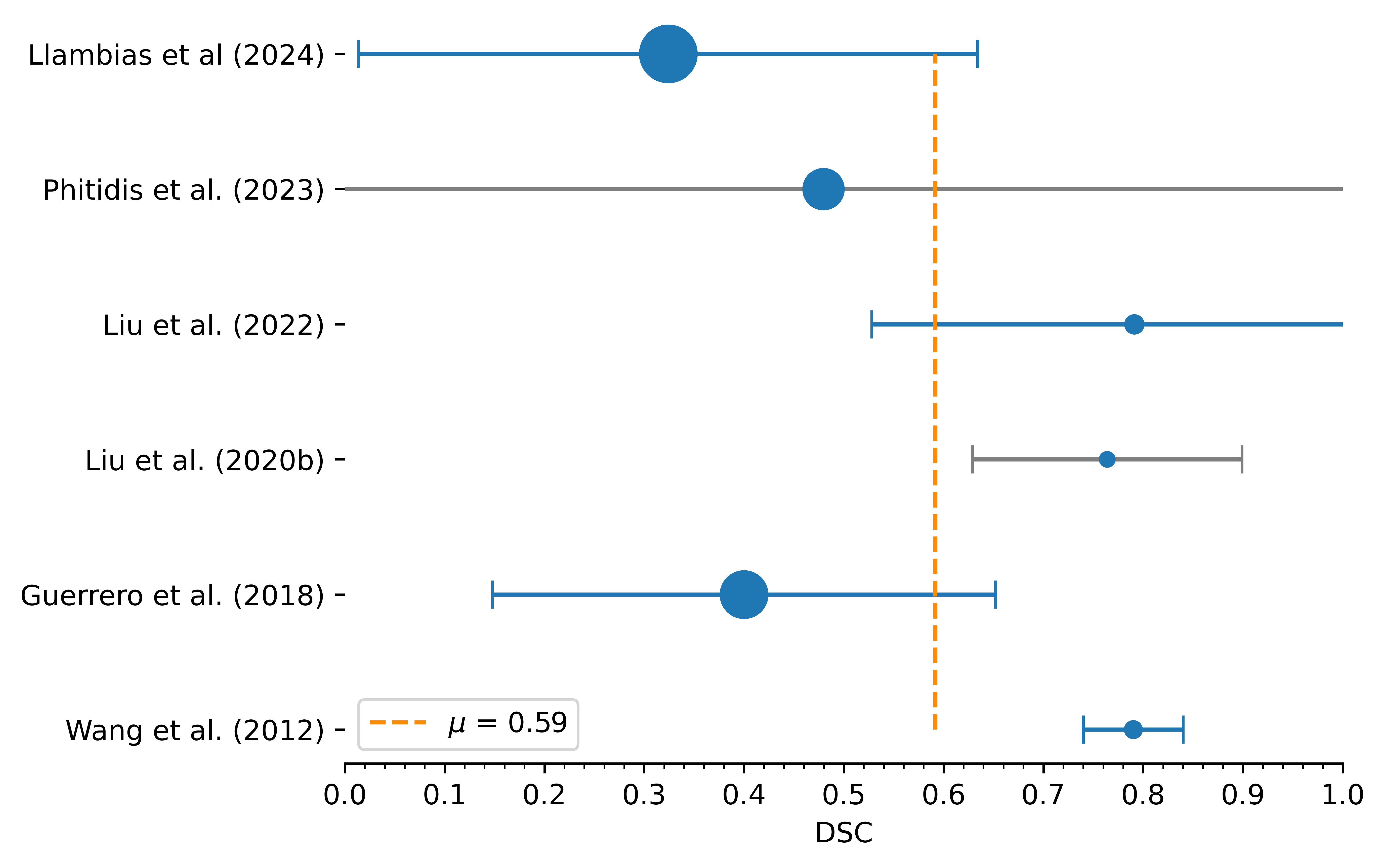}
        \caption{ISL}
        \label{fig:forest ISL research}
    \end{subfigure}
    \caption{Reported Dice similarity coefficients (DSC) for WMH and ISL. Error bars are standard deviation and marker size corresponds to test dataset size. Grey error bars indicated that the standard deviation was not reported and has been estimated visually from box plots as 3/4 of the interquartile range (assumes normal distribution).}
    \label{fig:forest research}
\end{figure}

\paragraph{White matter hyperintensities} Of the 10 papers which perform WMH segmentation and provide a DSC, 9 provide the standard deviation of the DSC (or a box plot from which we have estimated it). These results are shown in \textbf{Fig \ref{fig:foreset WMH research}}, where it can be seen that the average reported mean value for this subset is 0.76 and the lowest is 0.691. Not included in the plot, \citet{duan_primary_2020} report 0.666. It should be emphasised that these studies used different datasets and so direct comparison of the scores achieved has limited value.

\paragraph{Ischaemic stroke lesions} Out of 9 papers tackling ISL segmentation, 7 stated a DSC and 5 presented the standard deviation of the DSC (or a box plot from which we have estimated it). The reported mean DSC in this subset averages 0.59, but there is a wide range in results (0.324 to 0.791), indicating unreliable performance, but also potentially variation caused by different task settings (subcortical or cortical strokes, presence of WMH, use of DWI, etc). \citet{duan_primary_2020}, which is not included in the plot, reported a mean DSC of 0.728 for subcortical ISL which they segment in DWI images. \citet{uchiyama_computer-aided_2008} did not report an overlap metric, but classify ISL versus PVS with an AUC of 0.945. However, they stated a specificity of only 0.75, which may be due to the class imbalance present in their training data (89 ISL versus 20 PVS instances). \citet{liu2021deep} provide an overlap metric for foreground (WMH or ISL) vs background, achieving a DSC of 0.742.

\paragraph{Enlarged perivascular spaces} Of the 2 papers considering PVS detection, neither provided the DSC. \citet{uchiyama_computer-aided_2008} did not provide any metrics for PVS, since their goal was to differentiate ISL from PVS. \citet{sudre_3d_2018} proposed a model for detecting small objects including lacunes and PVS, but after detecting some uncertainties in their test cases, focused the validation instead on the overall sensitivity of their model to these small objects (0.727) and the median predicted bounding box overlap (59\%).

\paragraph{Lacunes} Of the 2 methods concerned with the identification of lacunes, 1 provided their DSC to be 0.496 \citep{duan_primary_2020} whilst the other only provided small object sensitivity and bounding box overlap as described above \citep{sudre_3d_2018}.

\paragraph{Cerebral microbleeds} Only \citet{duan_primary_2020} attempt CMB segmentation; they use  T2* MRI and achieve a DSC of 0.503.

\subsubsection{Risk of bias and applicability}

The results of the analysis of the risk of bias and applicability can be seen in \textbf{Table \ref{tab:quadas}}.


\vspace{10pt}
\begin{table}[H]
\tiny
\centering
\caption{Identified research publications - high-level overview. }
\label{tab:papers1}
\resizebox{\textwidth}{!}{%

\normalsize

\begin{table}[!htb]
\centering
\caption{Risk of bias assessment of the studies reviewed using the QUADAS 2 tool.}
\label{tab:quadas}
\resizebox{\textwidth}{!}{%
%
 \\ \hline
Uchiyama et al 2008 &
  \multicolumn{1}{l|}{Medium} &
  \multicolumn{1}{l|}{Bad} &
  \multicolumn{1}{l|}{Medium} &
  \multicolumn{1}{l|}{Medium} &
  \multicolumn{1}{l|}{Good} &
  Good &
  \multicolumn{1}{l|}{Good} &
  \multicolumn{1}{l|}{Good} &
  Bad \\ \hline
Wang et al 2012 &
  \multicolumn{1}{l|}{Medium} &
  \multicolumn{1}{l|}{Good} &
  \multicolumn{1}{l|}{Good} &
  \multicolumn{1}{l|}{Bad} &
  \multicolumn{1}{l|}{Good} &
  Bad &
  \multicolumn{1}{l|}{Good} &
  \multicolumn{1}{l|}{Good} &
  Good \\ \hline
Shi et al 2013 &
  \multicolumn{1}{l|}{Medium} &
  \multicolumn{1}{l|}{Good} &
  \multicolumn{1}{l|}{Good} &
  \multicolumn{1}{l|}{Bad} &
  \multicolumn{1}{l|}{Good} &
  Bad &
  \multicolumn{1}{l|}{Good} &
  \multicolumn{1}{l|}{Good} &
  Good \\ \hline
Tsai et al 2014 &
  \multicolumn{1}{l|}{Good} &
  \multicolumn{1}{l|}{Bad} &
  \multicolumn{1}{l|}{Bad} &
  \multicolumn{1}{l|}{Medium} &
  \multicolumn{1}{l|}{Good} &
  Bad &
  \multicolumn{1}{l|}{Good} &
  \multicolumn{1}{l|}{Good} &
  Good \\ \hline
Guererro et al 2018 &
  \multicolumn{1}{l|}{Good} &
  \multicolumn{1}{l|}{Good} &
  \multicolumn{1}{l|}{Medium} &
  \multicolumn{1}{l|}{Good} &
  \multicolumn{1}{l|}{Medium} &
  Bad &
  \multicolumn{1}{l|}{Good} &
  \multicolumn{1}{l|}{Good} &
  Good \\ \hline
Sudre et al 2019 &
  \multicolumn{1}{l|}{Good} &
  \multicolumn{1}{l|}{Good} &
  \multicolumn{1}{l|}{Good} &
  \multicolumn{1}{l|}{Medium} &
  \multicolumn{1}{l|}{Good} &
  Good &
  \multicolumn{1}{l|}{Good} &
  \multicolumn{1}{l|}{Good} &
  Good \\ \hline
Duan et al 2020 &
  \multicolumn{1}{l|}{Good} &
  \multicolumn{1}{l|}{Bad} &
  \multicolumn{1}{l|}{Good} &
  \multicolumn{1}{l|}{Good} &
  \multicolumn{1}{l|}{Medium} &
  Good &
  \multicolumn{1}{l|}{Good} &
  \multicolumn{1}{l|}{Good} &
  Bad \\ \hline
Liu et al 2020 (a) &
  \multicolumn{1}{l|}{Good} &
  \multicolumn{1}{l|}{Good} &
  \multicolumn{1}{l|}{Good} &
  \multicolumn{1}{l|}{Medium} &
  \multicolumn{1}{l|}{Medium} &
  Good &
  \multicolumn{1}{l|}{Good} &
  \multicolumn{1}{l|}{Good} &
  Bad \\ \hline
Liu et al 2020 (b) &
  \multicolumn{1}{l|}{Good} &
  \multicolumn{1}{l|}{Good} &
  \multicolumn{1}{l|}{Good} &
  \multicolumn{1}{l|}{Medium} &
  \multicolumn{1}{l|}{Good} &
  Good &
  \multicolumn{1}{l|}{Good} &
  \multicolumn{1}{l|}{Good} &
  Bad \\ \hline
Liu et al 2021 &
  \multicolumn{1}{l|}{Good} &
  \multicolumn{1}{l|}{Bad} &
  \multicolumn{1}{l|}{Good} &
  \multicolumn{1}{l|}{OK} &
  \multicolumn{1}{l|}{Good} &
  Good &
  \multicolumn{1}{l|}{OK} &
  \multicolumn{1}{l|}{OK} &
  Good \\ \hline
Liu et al 2022 &
  \multicolumn{1}{l|}{Bad} &
  \multicolumn{1}{l|}{Good} &
  \multicolumn{1}{l|}{OK} &
  \multicolumn{1}{l|}{OK} &
  \multicolumn{1}{l|}{Good} &
  Bad &
  \multicolumn{1}{l|}{OK} &
  \multicolumn{1}{l|}{Good} &
  Bad \\ \hline
Phitidis et al 2023 &
  \multicolumn{1}{l|}{Good} &
  \multicolumn{1}{l|}{Good} &
  \multicolumn{1}{l|}{OK} &
  \multicolumn{1}{l|}{Good} &
  \multicolumn{1}{l|}{OK} &
  Good &
  \multicolumn{1}{l|}{Good} &
  \multicolumn{1}{l|}{Good} &
  Good \\ \hline
Llambias et al 2024 &
  \multicolumn{1}{l|}{Good} &
  \multicolumn{1}{l|}{Good} &
  \multicolumn{1}{l|}{Good} &
  \multicolumn{1}{l|}{Good} &
  \multicolumn{1}{l|}{OK} &
  Good &
  \multicolumn{1}{l|}{Good} &
  \multicolumn{1}{l|}{Good} &
  OK \\ \hline
\end{tabular}%
}
\end{table}

\section{Discussion}

\paragraph{Preponderance of solutions for single rather than multiple CVD markers} Many solutions for CVD marker identification were excluded due to our inclusion criteria of requiring \emph{multiple} CVD markers to be tackled. For instance, all research on ICH and/or LVO was excluded from this review (e.g.~the work by \citet{wang2021deep} for the RSNA ICH detection kaggle challenge\footnote{\url{https://www.kaggle.com/c/rsna-intracranial-hemorrhage-detection}} \citep{flanders_construction_2020}). Many of these research methods were stimulated by challenges such as ISLES which focus on single pathologies. It would be interesting to see these methods applied to the more holistic CVD assessment that we consider in this review. Unfortunately, research in this area has been hampered by a lack of publicly available datasets with multiple CVD markers present and labelled by experts. Further, with the exception of Enterprise CTB, no other commercial systems claim to identify PVS or CMB and none identify lacunes. These are important markers for the assessment and management of CVD and the ability to automated their detection and quantification would provide a great benefit.

\paragraph{Vulnerability of commercial solutions to chronic ISL} WMH and subacute or chronic ISL have similar appearance in most MRI sequences; they can be difficult to differentiate and may co-occur in the same patient. Only two commercial solution jointly considered the detection of WMH and ISL. One is icobrain, although this is via independent subsystems intended for use in different situations: an acute stroke segmentation system requiring CTP; and a WMH segmentation system requiring T1-weighted and FLAIR MRI sequences. The other is uAI Discover CSVD, although sufficient information was not found for this product; other than this, no commercial systems considered chronic ISL. There were various research publications tackling this difficult problem.

\paragraph{Lack of publicly available data for CVD markers} We already mentioned that most public datasets focus on a single pathology e.g.~ISLES or the RSNA ICH detection kaggle challenge. To the best of our knowledge, until the recent ``Where is VALDO'' challenge \citep{sudre_where_2024}, there were no public datasets with annotations for lacunes, PVS or CMB, even individually. This holds back progress toward the development of a complete CVD support system and points to a lack of awareness surrounding the importance of these imaging markers in CVD assessment, compared to WMH, ISL, and ICH.

\paragraph{Lack of commercial solutions for MRI in acute stroke} It is unsurprising that the acute stroke systems almost all use NCCT, since thanks to its speed and accessibility, it is the most widely used imaging modality. However, DWI has been shown to be more sensitive to acute ischaemic changes \citep{lansberg_comparison_2000}. Therefore, DWI may see more uptake in the future and so systems to identify ISL using DWI may be highly desirable if this does happen. For the general neuroradiology support systems on the other hand, we see MRI being used exclusively. T1-weighted is mostly used for brain region segmentation due to its high level of detail and contrast between tissues, while FLAIR is most often utilised for identifying WMH, because it offers superior contrast between these features and surrounding tissue.

\paragraph{Bias towards male subjects} Of the 5 research papers which reported the sex of subjects, 4 used data with a bias towards male subjects, which may have introduced bias in the algorithms. Additional evaluation on female patients would be valuable to understand the performance of the proposed methods. It should be noted that these are small sample sizes and hence finding biases is expected.

\paragraph{Lack of methodological transparency and reproducibility} In terms of risk of bias, the most entries labelled as ``bad'' in \textbf{Table \ref{tab:quadas}} were in the columns headed ``Were all subjects included?'' and ``Are there concerns about reproducibility?''. These were mainly due to unexplained filtering of the available data and insufficient detail to reproduce the method, respectively. A lack of standardised benchmark datasets renders the meaningful comparison of metrics between methods impossible, further inhibiting transparency and reproducibility.

\paragraph{Popularity of convolutional neural networks} Of the subsystems with information on the method, 95\% of commercial stroke subsystems and 62\% of general commercial neuroradiology support subsystems employ DL. While transformers \citep{vaswani_attention_2017} are the most widely used architecture in natural language processing today, CNNs still dominate medical image analysis. This may be because their inductive biases (spatial invariance and locality) - while limiting their potential to model global context - allow them to learn more readily from limited data, which is important in the medical domain. Of the DL methods identified in this review, only one single product utilises a VIT architecture (Enterprise CTB); this is facilitated by a training dataset of 200,000 images. 

The oldest four research publications use rule-based methods and/or classical machine learning techniques, while the most recent nine all use DL. This highlights a shift in researchers' methods of choice, likely driven by the success of DL in other computer vision applications. All DL based methods identified in this reviews use variations of CNNs, mainly with a U-Net architecture. Again, due to the lack of standardised benchmark datasets, it is not currently possible to say whether this transition to DL has benefited algorithms' performance.

In general, DL provides some obvious benefits as well as some clear drawbacks when compared to other methods. Firstly, once trained, inference is usually very fast. For example, the publicly available DL model SynthSeg \citep{billot2023synthseg} can perform brain anatomy segmentation in under a minute, whereas the popular Freesurfer tool fits a model for each new images, meaning that it can take several hours to segment a single scan. ATLAS propagation based segmentation methods also require model fitting at test time to find the optimal transformation. These differences may have a significant impact on product viability, since brain anatomy segmentation is a key stage in many of the identified systems, especially those quantifying brain atrophy. While inference speed is a benefit of using DL models, the hardware required to run them efficiently (GPUs with tens of gigabytes of memory) is expensive. Additionally, training them often requires several days or even weeks of continuous energy usage, which is not environmentally friendly. DL models usually have millions of learnable parameters, which is what gives them their ability to model complex functions. However, this means that they are data hungry, and so large amounts of data are required to train them effectively. Their design also limits their explainability (i.e.~our ability to step through and understand the impact of each stage of the computation on the final prediction). Classical method such as random forests have far less parameters and so they are less prone to overfit the training data when it is limited. We are also able to examine and intuitively understand the nodes of each decision tree. All machine learning based methods depend heavily on the quality and coverage of the training data.

\paragraph{Integration into the clinical workflow} A significant barrier to the adoption of new technologies into the clinical workflow is the potential complexity of systems integration. Whilst this is undoubtedly a challenge, several of the commercial systems identified in this review have shown that it is not an insurmountable one, and that the positive clinical impact of the investment can be substantial (see the ``Impact'' column in \textbf{Table 1} of the supplementary material).

\paragraph{Limitations} Data extraction was performed by a single author (J.P.) following the search strategy. Commercially available software systems are constantly evolving with technology and consumer demand, meaning that the details presented in this systematic review may become outdated.

\section{Conclusion}
There is at this time no comprehensive, fully automated and well validated neuroradiological support system for all non-acute CVD findings (ISL, WMH, PVS, CMB, lacunes, and atrophy). We propose that such a system would provide a great benefit to society, as CVD is a leading cause of dementia and stroke, and automatic quantification of all of these features would open the door for more large-scale research and more efficient patient management. This could build on the success of acute stroke detection systems which are currently being utilised in several hospitals around the world with a reportedly positive impact on patient care, and software used in clinics to automate the arduous manual process of brain volume estimation and WMH segmentation.

The recent advancements in DL hold promise for the future of automatic evaluation of CVD markers. Further methodological innovations in the areas of data-efficient learning and small lesion segmentation, as well as an increase in the availability of standardised benchmark datasets and metrics (using best practices agreed by the research community \citep{maier2024metrics,reinke2024understanding}), will likely be required before development of a complete and robust system is feasible.

\section*{CRediT authorship contribution statement}
\noindent 
\textbf{Jesse Phitidis}: Conceptualization, Methodology, Data curation, Writing - original draft.
\textbf{Alison Q O'Neil}: Writing - original draft, Writing - review \& editing, Supervision.
\textbf{William N Whiteley}: Writing - review \& editing, Resources, Supervision.
\textbf{Beatrice Alex}: Writing - review \& editing, supervision.
\textbf{Joanna M. Wardlaw}: Writing - review \& editing, Resource, Supervision.
\textbf{Miguel O. Bernabeu}: Writing - review \& editing, Supervision.
\textbf{Maria Vald\'{e}s Hern\'{a}ndez}: Conceptualization, Methodology, Resources, Writing - original daft, Writing - review \& editing, Supervision.

\section*{Declaration of competing interests}
\noindent Alison Q O'Neil and Jesse Phitidis are employed (Alison Q O'Neil) or funded (Jesse Phitidis) by Canon Medical Research Europe - an organisation selling commercial medical imaging software. Canon Medical Research Europe has a partnership with Avicenna.AI.

\section*{Acknowledgements}  
\noindent 
Funding: Medical Research Scotland [ref. PHD-50441-2021]; Canon Medical Research Europe; BHF Data Science Centre (at Health Data Research UK); National Institute of Health research (UK); Alzheimer's Disease Data Initiative; the Neurii initiative which is a partnership among Eisai Co., Ltd, Gates Ventures, LifeArc and HDR UK; Alzheimer's Society; The Stroke Association; Legal \& General Group via Advanced Care Research Centre at University of Edinburgh; NIHR via the Artificial Intelligence and Multimorbidity: Clustering in Individuals, Space and Clinical Context (AIM-CISC) project [NIHR202639]; The Row Fogo Charitable Trust [Ref No: AD.ROW4.35. BRO-D.FID3668413]; Fondation Leducq Perivascular Spaces in Small Vessel Disease [16 CVD 05]; UK Dementia Research Institute [award no. UKDRI – Edin002, DRIEdi17/18, and MRC MC\_PC\_17113]  which receives its funding from DRI Ltd, funded by the UK Medical Research Council, Alzheimer’s Society and Alzheimer’s Research UK; Stroke Association/BHF/Alzheimer’s Society ‘Rates Risks and Routes to Reduce Vascular Dementia’ (R4VaD) Priority Programme Award in Vascular Dementia [16 VAD 07] ; British Heart Foundation Centre for Research Excellence Award III [RE/18/5/34216]; British Heart Foundation and The Alan Turing Institute Cardiovascular Data Science Award [C-10180357]; Dementias Platform UK 2 – Integrated Dementia Experimental Medicine. UK Medical Research Council; [MR/T033371/1]; The Galen and Hilary Weston Foundation [ref UB190097]; Fondation Leducq Transatlantic Network of Excellence [17 CVD 03]; EPSRC [grant no. EP/X025705/1]; Diabetes UK [20/0006221]; Fight for Sight [5137/5138]; the SCONe projects funded by Chief Scientist Office, Edinburgh \& Lothians Health Foundation, Sight Scotland, the Royal College of Surgeons of Edinburgh, the RS Macdonald Charitable Trust, and Fight For Sight.


\begin{landscape}

\appendix
\section{Supplementary Material}

\noindent For commercial systems, detailed information on the methods and training data were mostly not available through the websites, whitepapers, explicitly cited publications, or FDA premarket notifications. Since this is a particular point of interest, we additionally looked at: a) the linked publications on the website; and b) the first page of results for a Google Scholar search of the form \textit{company AND sub-product AND pathology}. It should be noted that as commercial software, algorithms and training data are likely to continuously evolve and so where possible, the most recent information is referenced.

\clearpage
\tiny
\begin{longtable}[c]{lllllll}
\caption{Commercial systems - details. With the exception of methods and training data (which were identified via linked publications or independent search where required and possible), the entries in this table are from the company website, whitepapers, explicitly cited publications, or FDA premarket notifications. Where these have not provided information, this has been stated, and where there are multiple sources of validation results, the most detailed has been presented. Linked publications are publications linked on the company website or in brochures, but not necessarily explicitly cited.}
\label{tab:commercial2}\\
\hline
\multicolumn{1}{|l|}{\textbf{Product}} &
  \multicolumn{1}{l|}{\textbf{\begin{tabular}[c]{@{}l@{}}(sub-product)\\ Population\end{tabular}}} &
  \multicolumn{1}{l|}{\textbf{\begin{tabular}[c]{@{}l@{}}(Sub-product)\\ Reference standard\end{tabular}}} &
  \multicolumn{1}{l|}{\textbf{\begin{tabular}[c]{@{}l@{}}(sub-product)\\ Method/Training\end{tabular}}} &
  \multicolumn{1}{l|}{\textbf{\begin{tabular}[c]{@{}l@{}}(sub-product)\\ Validation\end{tabular}}} &
  \multicolumn{1}{l|}{\textbf{\begin{tabular}[c]{@{}l@{}}(sub-product)\\ Impact\end{tabular}}} &
  \multicolumn{1}{l|}{\textbf{\begin{tabular}[c]{@{}l@{}}Linked\\ publications\end{tabular}}} \\ \hline
\endfirsthead
\multicolumn{7}{c}%
{{\bfseries Table \thetable\ continued from previous page}} \\
\endhead
\multicolumn{1}{|l|}{\begin{tabular}[c]{@{}l@{}}Always-on \\ AI\end{tabular}} &
  \multicolumn{1}{l|}{\begin{tabular}[c]{@{}l@{}}(BriefCase ICH)\\ Training:\\ $\sim$50,000 cases from 9 sites\\ and 17 different scanners \\ (63).\\ Validation:\\ 3 sites (2 in USA), 198 cases\\  ($\sim$50\% with ICH) (1).\\ \\ (BriefCase LVO)\\ Validation:\\ 3 sites in USA, 383 cases \\ with LVO (2).\\ \\ (BriefCase BA)\\ Validation:\\ 5 sites, multinational, 286 \\ cases (96 with BA) (3).\\ \\ (icobrain-cva)\\ Validation:\\ A clinical dataset and digital \\ phantom (4).\end{tabular}} &
  \multicolumn{1}{l|}{\begin{tabular}[c]{@{}l@{}}(BriefCase BA)\\ Validation:\\ Ground thruth generated by \\ 2 US board-certified radio-\\ logists with a third in case \\ of disagreement (3).\end{tabular}} &
  \multicolumn{1}{l|}{\begin{tabular}[c]{@{}l@{}}(BriefCase ICH)\\ 3D CNN proposes regions\\ and second stage filters \\ based on final layer \\ features (64).\\ \\ (BriefCase LVO)\\ CNN (65).\\ \\ (icobrain-cva)\\ CNN trained to predict\\ AIF by optimising lesion\\ segmentation objective \\ (7).\end{tabular}} &
  \multicolumn{1}{l|}{\begin{tabular}[c]{@{}l@{}}(BriefCase ICH)\\ Sen = 0.936 (0.866-0.976),\\ Spe = 0.923 (0.854-0.966) \\ (1).\\ \\ (BriefCase LVO)\\ Sen = 0.888 (0.819-0.938),\\ Spe = 0.872 (0.825-0.911) \\ (2).\\ \\ (BriefCase BA)\\ Sen = 0.885 (0.804-0.941),\\ Spe = 0.895 (0.84-0.937) \\ (3).\end{tabular}} &
  \multicolumn{1}{l|}{\begin{tabular}[c]{@{}l@{}}(BriefCase ICH)\\ In 59 TP cases the mean time-\\ to-notification (4.46 min) was \\ 68.11 min faster than mean \\ time-to-exam-open by \\ radiologist (1).\\ \\ (BriefCase LVO)\\ In 111 cases the mean time-to-\\ notification was 3.8 min (2).\\ \\ (BriefCase BA)\\ In 65 TP cases in 2 sites in \\ USA the mean time-to-noti-\\ fication (4.2 min) was 85.2 \\ min faster than mean time-to-\\ exam-open by radiologist (3).\end{tabular}} &
  \multicolumn{1}{l|}{\begin{tabular}[c]{@{}l@{}}(aidoc)\\ (5)\\ \\ (icobrain-cva)\\ (6, 7)\end{tabular}} \\
\end{longtable}
\end{landscape}

\bibliographystyle{elsarticle-harv} 
\bibliography{references.bib}





\end{document}